\documentclass[prb,a4paper,showpacs,twocolumn]{revtex4}

\usepackage{amssymb}
\usepackage{amsmath}
\usepackage{amsfonts}
\usepackage{graphicx}
\usepackage{bm}
\usepackage{color}

\renewcommand{\Re}{\,\textrm{Re}\,}
\renewcommand{\Im}{\,\textrm{Im}\,}
 
\DeclareMathOperator{\tr}{tr} \DeclareMathOperator{\sign}{sgn}

\definecolor{DarkBlue}{rgb}{0,0,0.6}

\font\tencal =rsfs10 scaled \magstep 1
\def\CD{\hbox{\tencal D}}

\begin{document}

\title{Charge relaxation resistance in the Coulomb blockade problem}

\author{Ya.I.\,Rodionov$^{1}$, I.S.\,Burmistrov$^{1,2}$ and
A.S.\,Ioselevich$^{1}$}

\affiliation{$^{1}$ L.D.\ Landau Institute for Theoretical Physics,
Russian Academy of Sciences, 117940 Moscow, Russia}
\affiliation{$^{2}$ Department of Theoretical Physics, Moscow
Institute of Physics and Technology, 141700 Moscow, Russia}

\begin{abstract}
We study the dissipation in a system consisting of a small metallic
island coupled to a gate electrode and to a massive reservoir via
single tunneling junction. The dissipation of energy is caused by a
slowly oscillating gate voltage. We compute it in the regimes of
weak and strong Coulomb blockade. We focus on the regime of not very
low temperatures when electron coherence can be neglected but
quantum fluctuations of charge are strong due to Coulomb
interaction. The answers assume a particularly transparent form
while expressed in terms of specially chosen physical observables.
We discovered that the dissipation rate is given by a universal
expression in both limiting cases.
\end{abstract}
\date{\today}


\maketitle

%%%%%%%%%%%%%%%%%%%%%%%%%%%%%%%%%%%%%%%%%%%%%%%%%%%%%%%%%%%%%%%%%%%%%%%%%%%%%%%%
%
\section{Introduction\label{Sec:Intro}}
The phenomenon of Coulomb blockade has become an excellent tool for
observation of interaction effects in single electron devices.
Theoretical means for its exploration are well developed and
versatile.~\cite{zaikin,ZPhys,grabert,blanter,aleiner,Glazman} The
simplest mesoscopic system displaying Coulomb blockade is a single
electron box (SEB). The properties of such a system are essentially
affected by electron coherence and interaction. Our work is
motivated by a considerable recent theoretical and experimental
interest in the relation between dissipation and resistance of this
device in various parametric
regimes.~\cite{buttiker0,buttiker3,buttiker2,buttiker1,gabelli,delsing}
\begin{figure}[t]
  \includegraphics[width=70mm]{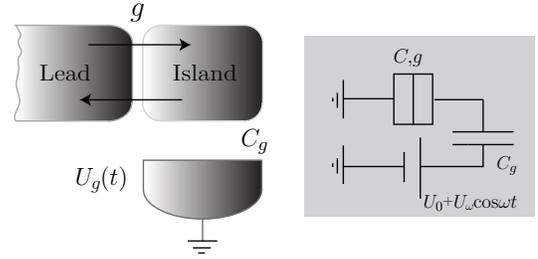}
  \caption{\label{figure1}
    Measurement of resistance $R_q$. The SEB is subjected to a constant gate voltage\ $U_0$
    The dissipative current through the tunneling contact is caused by a weak
    AC voltage\ $U(t)$.
          }
\end{figure}

The set-up is as follows (see Fig.\ref{figure1}). Metallic island is
coupled to an equilibrium electron reservoir via tunneling junction.
The island is also coupled capacitively to the gate electrode. The
potential of the island is controlled by the voltage\ $U_g$\ of the
gate electrode. The physics of the system is governed by several
energy scales: the Thouless energy of an island\ $E_{\rm Th}$, the
charging energy\ $E_c$, and the mean level spacing\ $\delta$.
Throughout the paper the Thouless energy is considered to be the
largest scale in the problem. This allows us to treat the metallic
island as a zero dimensional object with vanishing internal
resistance. The dimensionless conductance of a tunneling junction\
$g$\ is an additional control parameter.

Initially, the main quantity of interest in a Coulomb blockaded SEB
was its effective capacitance:\ $\partial Q/\partial U_g$, where\
$Q$\ is the average charge of a
island.~\cite{matveev,Grabert0,matveev1,Grabert1,Grabert2,beloborodov1}
Paper~[\onlinecite{buttiker0}] however sparked both theoretical and
experimental attention to the dynamic response functions of such a
set-up.~\cite{buttiker3,buttiker2,buttiker1,imry,Park,gabelli,delsing}
It is worthwhile to mention that the system does not allow for
conductance measurements since there is no DC-transport. This way an
essential dynamic characteristic becomes the set-up admittance,
which is a current response to an AC-gate voltage
$U_g(t)=U_0+U_\omega \cos\omega t$. As it is well-known, the real
part of admittance determines energy dissipation in an electric
circuit. Classically, the average energy dissipation rate of a
single electron box is given as follows
\begin{gather}
 \label{admittance0}
\mathcal{W}_\omega=\omega ^2C^2_g R|U_\omega|^2,\qquad
R=\frac{h}{e^2 g},\ \qquad
 \hbar \omega\ll gE_c,
\end{gather}
where $C_g$ denotes the gate capacitance, $e$ - the electron charge,
and $h=2\pi \hbar$ - the Planck constant. Expression
\eqref{admittance0} presents us with a natural way of extracting the
resistance of a system from its dissipation power. The resistance of
a classical system is thus fully determined by the tunneling
conductance of the contact via Kirchhoff's law:\ $R=h/(e^2 g)$. The
question one asks is how quantum effects such as electron coherence
and interaction change this result? One expects that correct quantum
dissipation is going to give generalized quantum resistance. The
obvious stumbling block one foresees is that only combination of two
observables:\ $C_g^2R$\ can be extracted from the dissipation power
rather than just\ $R$. For the case of fully coherent SEB this key
difficulty was resolved in Ref.~[\onlinecite{buttiker0}]. It was
shown that the energy dissipation rate $\mathcal{W}_\omega$ can be
factorized in accordance with its classical appearance
\eqref{admittance0} but the definition of physical quantities
comprising it becomes different. Geometrical capacitance\ $C_g$\
should be substituted by a new observable: mesoscopic capacitance\
$C_\mu$. This leads to the establishment of another observable: {\it
charge relaxation} resistance\ $R_q$ such that \ $R\rightarrow R_q$
in Eq.~\eqref{admittance0}. Charge relaxation resistance of a
coherent system differs drastically from its classical counterpart.
In particular, as shown in Ref.~[\onlinecite{buttiker0}], the charge
relaxation resistance of a single channel junction doesn't depend on
its transmission.
%It was measured in the recent experiment by Gabelli {\it et al} \cite{gabelli}.
The admittance in the quasi-static regime was investigated
% the focal point of
in the recent experiment by Gabelli et al.~\cite{gabelli} The
measurements were performed at low temperatures\ $T\lesssim \delta$\
when the system could be regarded as coherent. The question that has
remained unattended by the theory is what happens to dissipation and
resistance at transient temperatures when thermal fluctuations smear
out electron coherence but electron-electron interaction is strong?
The recent experiment by Persson et al~\cite{delsing} explored the
energy dissipation rate at these transient temperatures.

Motivated by the experiment [\onlinecite{delsing}] we address the
same question from the theoretical point of view. We study the
energy dissipation rate of a single electron box in the so-called
`interactions without coherence' regime.  It corresponds to the
following hierarchy of energy scales:\ $E_{\rm Th}\gg E_c\gg
T\gg\max\{\delta, g\delta\}$. This temperature regime is such that
keeps electrons strongly correlated $(T\ll E_c)$, yet allows to
discard electron coherence\ $(T\gg \max\{\delta,
g\delta\})$.~\cite{beloborodov,efetov} We compute the energy
dissipation rate and the SEB admittance in the limits of large\
$(g\gg1)$\ and small\ $(g\ll1)$\ dimensionless tunneling conductance
of the junction.

We consider a multichannel junction but the conductance of each
channel is assumed to be small\ $g_{\rm ch}\ll1$. Then, the physics
of the system is most adequately described in the framework of
Ambegaokar-Eckern-Sch\"{o}n (AES) effective
action.~\cite{ambegaokar} Our results lead to the generalization of
classical result \eqref{admittance0}. We found that at $\omega\to 0$
the average energy dissipation rate can be factorized in both\
$g\gg1$\ and\ $g\ll1$\ limits as
\begin{gather}
   \label{eq1}
   \mathcal{W}_\omega=\omega^2C^2_g(T)R_q(T)|U_\omega|^2, \qquad R_q(T)=\frac{h}{e^2 g^\prime(T)},
\end{gather}
in complete analogy with classical expression \eqref{admittance0}.
Here,\ $R_q(T)$ and\ $C_g(T)$\ are identified as charge relaxation
resistance and renormalized gate capacitance, respectively. It is
worthwhile to mention that the physical observables \ $g^\prime(T)$\
and\ $C_g(T)$\ are defined universally for any value of
dimensionless conductance\ $g$. It allows us to suggest that
Eq.~\eqref{eq1} remains valid for arbitrary value of $g$.

\begin{figure}%[t]
  \includegraphics[width=88mm]{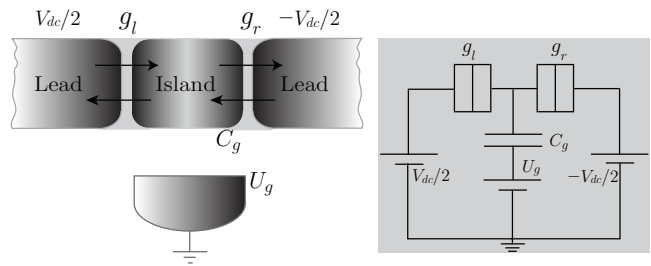}
  \caption{\label{figure2}
   Measurement of conductance. The SET is subjected to a constant gate voltage\ $U_g$
    and constant bias\ $U$.
          }
\end{figure}

In order to explain physics behind quantities\ $g^\prime(T)$\ and\
$C_g(T)$,\ it is useful to consider a single electron transistor
(SET) rather than SEB (see Fig.~\ref{figure2}).
In the absence of DC-voltage between left and right reservoirs a SET
represents essentially a SEB except different definition of the
parameter $g$. Then,\ $g^\prime(T)$ is the very quantity that
determines the SET conductance. The renormalized gate capacitance\
$C_g(T)$ is very different from the effective capacitance $\partial
Q/\partial U_0$. In fact,\ $C_g(T)=\partial q^\prime(T)/\partial
U_0$,  where $q^\prime(T)$\ is the physical observable introduced
recently in Ref.~[\onlinecite{burmistrov1}] to describe the\
$\theta$-angle renormalization in the Coulomb blockade problem. The
quantity $q^\prime$ is determined not only by the average charge $Q$
but also by the anti-symmetrized (so-called, quantum) current noise
in a SET.

The paper is organized as follows. Section II is used to introduce
AES model. Sections III and IV are devoted to dissipation in the
weak ($g\gg1$) and strong ($g\ll1$) coupling regimes. Section V is
devoted to discussion.

%
%%%%%%%%%%%%%%%%%%%%%%%%%%%%%%%%%%%%%%%%%%%%%%%%%%%%% The set up %%%%%%%%%%%%%%%%%%%%%%%%%%%%%%%%%%%%%%%%%%%%%%%%%%%%%%
%

\section{Formalism\label{Sec:First}}

\subsection{Hamiltonian\label{Sec:First:H}}

A single electron box is described by the Hamiltonian
\begin{gather}
   \label{ham1}
      H=H_0+H_c+H_t ,
\end{gather}
where\ $H_0$\ describes free electrons in the lead and the island,\
$H_c$\ describes Coulomb interaction of carriers in the island, and\
$H_t$\ describes the tunneling.
\begin{gather}
   \label{ham2}
    \begin{split}
    H_0&=\sum_{k}\varepsilon^{(a)}_{k}a^\dagger_{k}a_{k}+
    \sum_\alpha\varepsilon^{(d)}_\alpha d^\dagger_{\alpha}d_{\alpha}.\\
    \end{split}
\end{gather}
Here, operators $a^\dag_{k}$ ($d^\dag_{\alpha}$)\ create a carrier
in the lead (island).
\begin{gather}
   \label{ham-tun}
    H_{t}=\sum_{k,\alpha}t_{k\alpha} a_{k}^\dagger d_{\alpha}+{\rm
    h.c.}
\end{gather}
The charging Hamiltonian of electrons in the box is taken in the
capacitive form:
\begin{gather}
   \label{ham3}
      H_c=E_c\big(\hat{n}_d-q\big)^2
\end{gather}
Here,\ $E_c= e^2/(2C)$\ denotes the charging energy, and\
$q=C_gU_g/e$\ the gate charge. $\hat{n}_d$\ is an operator of a
particle number in the island:
\begin{gather}
   \label{number}
    \begin{split}
    \hat{n}_d=\sum_{\alpha}d^\dagger_{\alpha}d_{\alpha}.
    \end{split}
\end{gather}

It is convenient to introduce hermitean matrices:
\begin{gather}
   \hat{g}_{kk^\prime}=(2\pi)^2
   \left[\delta(\varepsilon^{(a)}_{k})\delta(\varepsilon^{(a)}_{k^\prime})\right]^{1/2}\sum_\alpha t_{k\alpha}
   \delta(\varepsilon^{(d)}_{\alpha})t^\dagger_{\alpha
   k^\prime},\\
   \hat{\tilde{g}}_{\alpha\alpha^\prime}=(2\pi)^2
   \left[\delta(\varepsilon^{(d)}_{\alpha})\delta(\varepsilon^{(d)}_{\alpha^\prime})
   \right]^{1/2}\sum_k t^\dagger_{\alpha
   k}
   \delta(\varepsilon^{(a)}_{k})t_{k\alpha^\prime},
\end{gather}
the first of them acting in the Hilbert space of the states of the
lead, the second -- in the space of the islands states. The energies
$\varepsilon^{(a)},\varepsilon^{(d)}$ are accounted for with respect
to the Fermi level, and the delta-functions should be smoothed on
the scale $\delta E$, such that $\delta\ll\delta E\ll T$.

The eigenstates of $\hat{g}$ ($\hat{\tilde{g}}$) describe the
`channel states' in the lead (island), while the transmittances of
the corresponding channels\ ${\cal T}_\gamma$\ are related to the
eigenvalues $g_{\gamma}$. Note, that in general the rank of the
matrix $\hat{g}$ differs from that of the matrix $\hat{\tilde{g}}$,
so that the numbers of eigenvalues are also different. This
difference is, however, irrelevant, since it stems from the ``closed
channels'' with $g_{\gamma}\approx 0$, i.e., the states strongly
localized either within the lead, or within the island. The
effective `channel conductance'\ $g_{\rm ch}$\ and the effective
number of open channels $N_{\rm ch}$ can be defined as
\cite{glazman}
\begin{gather}
   \label{aes-condition1}
  g_{\rm ch}=\frac{\tr(\hat{g}^2)}{\tr\hat{g}},\quad N_{\rm ch}=\frac{(\tr\hat{g})^2}{\tr(\hat{g}^2)}.
  \end{gather}
  In general case the effective action can be written as a sum
  of terms, proportional to $\tr(\hat{g}^k)$, over all integer $k$ (see
ref.[\onlinecite{glazman}]). The problem is considerably simplified
in the tunnel case, when
\begin{gather}
   \label{aes-condition2}
     g_{\rm ch}\ll 1.
  \end{gather}
In the present paper we assume this condition to be satisfied.
Then all terms with $k>1$ can be neglected, compared to  the
leading term with $k=1$ and the standard form of the AES action
can be easily reproduced. In particular, the classical
dimensionless conductance of the junction is expressed as:
\begin{gather}
    \label{conductance-def1}
   g=\tr\hat{g}=\tr\hat{\tilde{g}}=g_{\rm ch}N_{\rm ch}.
\end{gather}
Note, that under the condition \eqref{aes-condition2}  $g$  still
can be large, if  the number of channels $N_{\rm ch}\gg 1$ is
sufficiently large.

 Throughout the paper we keep the units such that\ $\hbar=e=1$,
except for the final results.

\subsection{Conductance and dissipation}
To study the electric properties of a system we compute energy
dissipation caused by slow oscillations of external gate voltage\
$U_g(t)=U_0+U_\omega\cos \omega t$.

The average energy dissipation rate can be found following the
standard scheme:~\cite{LLIII}
\begin{gather}
  \label{dissipation4}
   \mathcal{W}_\omega=\frac{d E}{dt}
   %=\frac{\partial E}{\d U_g}\frac{dU_g}{dt}
   =\Big\langle\frac{\delta H}{\delta
   U_g}\Big\rangle\frac{dU_g}{dt}.
\end{gather}
Here,\ $E$\ is the energy of the system,\ $H$ is given by
\eqref{ham1} and angular brackets denote full quantum statistical
average. Since
\begin{gather}
 \label{deriv1}
  \Big\langle\frac{\delta H}{\delta U_g}\Big\rangle=
  -\frac{C_g}{C}\sum_\alpha\langle d_\alpha^\dagger
  d_\alpha\rangle+\frac{C_g^2}{C}U_g,
\end{gather}
the energy dissipation is determined by a response of the electron
density in the island to the time-dependent gate voltage $U_g(t)$.
Therefore, it can be found via Callen-Welton fluctuation-dissipation
theorem:~\cite{LLV}
\begin{gather}
  \label{dissip-fdt}
 \mathcal{W}_\omega=\frac{C_g^2}{2 C^2}\omega \Im \Pi^R(\omega) |U_\omega|^2 .
\end{gather}
Here,\ $\Pi^R(\omega)$\ is the retarded electron polarization
operator:
\begin{gather}
  \label{polar-0}
 \Pi^R(t)=i \Theta(t) \langle[\hat n_d(t),\hat n_d(0)]\rangle,\ \
  \hat n_d=\sum_\alpha d_\alpha^\dagger d_\alpha
\end{gather}
with $\Theta(t)$ denoting Heaviside step function.

We are interested in the quasi-static regime $\omega\to 0$. Then, as
it will be proven below, the polarization operator\ $\Pi^R(\omega)$\
is possible to expand in regular series in $\omega$:
\begin{gather}
   \label{polar-3}
   \Pi^R(\omega)=\pi_0(T)+i\omega\pi_1(T)+{\cal O}(\omega^2) ,
\end{gather}
where both\ $\pi_0(T)$\ and\ $\pi_1(T)$\ are real functions of
temperature and other SEB parameters. Then the energy dissipation
rate is solely determined by the linear coefficient\ $\pi_1(T)$\ and
acquires Ohmic form:
\begin{gather}\label{dissgen}
  \mathcal{W}_\omega=\frac{\omega^2}{2}\mathcal{A}(T) |U_\omega |^2,\ \
  \mathcal{A}(T)=\frac{C_g^2}{C^2}\pi_1(T) .
\end{gather}

The SEB admittance $g(\omega)$ which is the linear reponse of an
AC-current $I_\omega$ to AC-gate voltage $U_\omega$:
$\mathcal{G}(\omega)=I_\omega/U_\omega$, is related to the
polarization operator (see Appendix):
\begin{gather}\label{admgen}
 \mathcal{G}(\omega)=-i\omega C_g\bigl (1+\Pi^R(\omega)/C\bigr ).
\end{gather}
As expected, the energy dissipation rate is proportional to the real
part of the admittance: $\mathcal{W}_\omega \sim \Re {\cal
G}(\omega)$. The static part of the polarization operator
$\Pi^R(\omega)$ is determined by the effective capacitance $\partial
Q/\partial U_0$ as
\begin{equation}
\pi_0(T) = \frac{C}{C_g} \frac{\partial Q}{\partial U_0} - C
,\label{SWId}
\end{equation}
where $Q=\langle \hat n_d \rangle$ denotes the average charge on the
island. We mention that Eq.~\eqref{SWId} is analogous to the
well-known Ward identity which relates static polarization operator
and compressibility.~\cite{AGD} Using
Eqs.~\eqref{polar-3}-\eqref{SWId}, we can establish the following
result:
\begin{gather}
 \label{admittance_q}
  \mathcal{G}(\omega)=-i\omega \frac{\partial Q}{\partial U_0}+\frac{C}{C_g}\mathcal{A}(T)\omega^2+{\cal O}(\omega^3)
\end{gather}
which is a quantum generalization of the classical relation
\begin{gather}
  \label{admittance_cl}
   \mathcal{G}(\omega)=-i\omega C_g+ C_g CR \omega^2+{\cal O}(\omega^3) .
\end{gather}
Therefore, both the admittance and the energy dissipation rate are
determined by the polarization operator $\Pi^R(\omega)$ which
involves one unknown function $\pi_1(T)$  in the quasi-static
regime.

\subsection{AES model\label{Sec:First:AES}}

The condition \eqref{aes-condition2} validates the use of
AES-effective action~\cite{ambegaokar} which describes the physics
of the set-up in terms of a single quantum phase\ $\varphi(\tau)$\
fluctuating in Matsubara time\ $\tau$:
\begin{gather}
    \label{action0}
    S_{AES}=S_d+S_g+S_c .
\end{gather}
Here,\ $S_d$\ is the dissipative part of the action in the standard
form:
\begin{gather}
  \label{action1}
 \begin{split}
 S_d&=-\frac{g}{4}\int_0^\beta\alpha(\tau_{12})e^{i\varphi(\tau_1)-i\varphi(\tau_2)}\,d\tau_1d\tau_2,\\
 \alpha(\tau)&=\frac{T^2}{\sin^2\pi
 T\tau}=-\frac{T}{\pi}\sum_{\omega_n}|\omega_n|e^{-i\omega_n\tau} ,
   \end{split}
\end{gather}
where\ $\beta=1/T$, $\tau_{12}=\tau_1-\tau_2$, $\omega_n =2\pi T n$
and $g$ is defined by \eqref{conductance-def1} and stands for the
dimensionless (in units $e^2/h$) conductance of the tunnel junction.
The term $S_g$\ represents a coupling with the gate voltage\ $U_0$:
\begin{gather}
  S_g=-iq \int_0^\beta \dot{\varphi} d\tau  = -2\pi q W i.
  %-iC_g\int_0^\beta{\dot\varphi}(\tau)U(\tau)d\tau,\ \ q=C_gU_0/e
 \label{action2}
\end{gather}
Here, integer $W$\  is the winding number of a field\
$\varphi(\tau)$ which appears through the constraint
\begin{equation}
\label{BoundCond} \varphi(\beta)-\varphi(0)=2\pi W.
\end{equation}
Non-zero value of $S_g$ appears for topologically non-trivial field
configurations only. The charging part of the action is as follows
\begin{gather}
   S_c=\frac{1}{4E_c}\int_0^\beta{\dot \varphi}^2\,d\tau.
  \label{action3}
\end{gather}
Physically, time derivative of the phase variable $\dot{\varphi}$
describes voltage fluctuations in a SEB. We emphasize that
AES-action is valid for any value of\ $g$. We work in the regime\
$T\ll E_c$. Charging term\ $S_c$\ is thus always small providing a
natural ultraviolet cut-off in the theory:\ $\Lambda=gE_c$.
%The second term in \eqref{action2} is a source term.
%It is proportional to AC gate voltage which causes dissipation in the system.

Our aim is to compute the polarization operator \eqref{polar-0}
which, according to Eqs~\eqref{dissgen} and \eqref{admgen},
determines the energy dissipation and admittance. Therefore, we need
to express initial observables cast in terms of fermionic operators
through correlators of bosonic field\ $\varphi(\tau)$. This is done
in part A of the appendix by employing Keldysh formalism.
%to avoid issues with cumbersome analytical continuation related to Matsubara basis.
The polarization operator $\Pi^R(\omega)$ then can be obtained by
analytical continuation $i\omega_n\to\omega+i0$ of the following
phase correlator in Matsubara basis:
\begin{gather}
  \label{dissipation2}
  \Pi(\tau)=-C^2\langle{\cal T}_\tau\dot{\varphi}(\tau)\dot{\varphi}(0)\rangle .
\end{gather}
Here ${\cal T}_\tau$ denotes time ordering. So far we made no
assumptions about the value of\ $g$. The AES model is however
impossible to tackle for arbitrary\ $g$s\ due to highly non-linear
form of the dissipative term.  In the next chapter of the paper we
restrict our attention to the case of large dimensionless
conductance\ $g\gg1$; the quantity\ $1/g$\ then becomes an expansion
parameter of perturbation theory.

%%%%%%%%%%%%%%%%%%%%%%%%%%%%%%%%%%%%%%%%%%%%%
%%%%%%%%%%%%%%%%%%%%%%%%%%%%%%%%%%%%%%%%%%%%%%
%%%%%%%%%%%%%%%%%%%%%%%%%%%%%%%%%%%%%%%%%%%%%%

\section{Weak coupling regime, $g\gg 1$\label{Sec:First2}}

\subsection{Perturbation theory\label{Sec:First2:PT}}
To expand the polarization operator\ $\Pi(i\omega_n)$\ in powers of
$1/g$ it is convenient to use the Matsubara frequency
representation,
\begin{equation}
\varphi(\tau) = \sum_{n} \varphi_n e^{-i\omega_n\tau},\quad
\varphi_{-n}=\varphi_n^*.
\end{equation}
Then, the quadratic part of AES action assumes the form:
\begin{equation}
S_{AES}^{(2)} = g \sum_{n>0}\left ( n + \frac{2\pi^2 T}{g E_c} n^2
\right ) |\varphi_n|^2 .\label{SAESquart}
\end{equation}
It determines the propagator of the $\varphi$ field as
\begin{equation}
\langle \varphi_n \varphi_{m}\rangle = \frac{1}{g}\,
\frac{\delta_{m,-n}}{|n|+2\pi^2 T n^2/(g E_c)} .\label{Prop}
\end{equation}
Evaluation of the polarization operator at the tree level yields
\begin{gather}
   \frac{\Pi(i\omega_n)}{C^2} =-\frac{2\pi|\omega_n|}{g}+\mathcal{O}(\omega_n^2).
\end{gather}
Performing standard one-loop calculations one finds
\begin{equation}
   \frac{\Pi(i\omega_n)}{C^2} =-\frac{2\pi|\omega_n|}{g}\left (1+ \frac{2}{g} \ln\frac{gE_c e^{\gamma+1}}{2\pi^2T} \right ) +
   \mathcal{O}(\omega_n^2).
\end{equation}
With the help of the renormalization group analysis this result can
be written as~\cite{hofstetter}
\begin{gather}
   \frac{\Pi(i\omega_n)}{C^2}=-\frac{2\pi|\omega_n|}{g(T)} +  \mathcal{O}(\omega_n^2). \label{PT-Pi}
\end{gather}
Here,\ $g(T)$\ is given by
\begin{gather}
   \label{conductance1}
   g(T)=g-2\ln\frac{gE_ce^{\gamma+1}}{2\pi^2T}
\end{gather}
with\ $\gamma\approx0.577$\ being Euler's constant.
Eq.~\eqref{conductance1} describes the well-known one-loop
temperature renormalization of the coupling constant.~\cite{perturb}

\subsection{Instantons\label{Sec:First2:Inst}}
So far the phenomenon of Coulomb blockade i.e. dependence on $q$,\
is completely absent in all our expressions for polarization
operator. To catch it we have to take into account instanton
solutions of AES action.~\cite{korshunov,Bulgadaev} Korshunov's
instantons read
\begin{gather}
  \label{inst1}
  e^{i\varphi_W(\tau |\{z_a\})}=\prod_{a=1}^{|W|}
  \bigg[\frac{e^{2\pi i\tau T}-z_a}{1-z^*_ae^{2\pi i\tau T}}\bigg]^{\sign W} .
\end{gather}
Here,\ $z_a$\ is a set of arbitrary complex numbers. Positive values
of winding numbers $W$ are assigned to instantons with $|z_a|<1$ and
negative ones to anti-instantons with  $|z_a|>1$. On the classical
solutions~\eqref{inst1} the dissipative\ $S_d$\ and topological\
$S_g$\ part of AES-action becomes
\begin{gather}
  \label{inst_action1}
  S_d[\varphi_W]+S_g[\varphi_W]=\frac{g}{2}|W|-2\pi W q i .
\end{gather}
It is finite and independent of\ $z_a$s. These parameters are
zero-modes. The charging term though does depend on them:
\begin{gather}
  S_c[\varphi_W]=\frac{\pi^2T}{E_c}\sum_{a,b}\frac{1+z_a z^*_b}{1-z_a z^*_b} .
\end{gather}
Thus\ $z_a$s can only be viewed as approximate zero modes and the
instanton configurations with $|z_a|\to 1$ are supressed.

As it is clear from Eq.~\eqref{inst_action1} every instanton brings
a small factor\ $e^{-g/2}$\ to any observable we want to compute. In
what follows, we restrict ourselves to one-instanton ($W=\pm1$)
contribution only.

\subsection{Instanton correction to the polarization operator\label{Sec:First2:InstPO}}

To get the instanton contribution to the polarization operator we
need to compute one-instanton correction to the correlator\
$\langle{\cal T}_\tau\dot{\varphi}(\tau)\dot{\varphi}(0)\rangle$. Up
to the one-instanton contributions we find
\begin{gather}
  -\frac{\Pi(i\omega_n)}{C^2}\simeq \langle\dot{\varphi}\dot{\varphi}\rangle^{(0)}_{\omega_n}\Big(1-\sum_{W=\pm1}\frac{{\cal Z}_W}{{\cal Z}_0}\Big)+
   \sum_{W=\pm1}\langle\dot{\varphi}\dot{\varphi}\rangle^{(W)}_{\omega_n} \notag  \\
  =\hbox{I}+\hbox{II}, \label{PiInst1}
\end{gather}
where $\langle \dot{\varphi}\dot{\varphi}\rangle_{\omega_n} =
\int_0^\beta \langle\dot{\varphi}(\tau)\dot{\varphi}(0)\rangle
\exp(i\omega_n\tau) d\tau$ and
\begin{gather}
\begin{split}
{\cal Z}_W &= \int_{W} \hbox{\tencal D}\,\varphi \exp[-S_{AES}], \\
\langle\dot{\varphi}\dot{\varphi}\rangle^{(W)} &= \frac{1}{{\cal
Z}_0} \int_{W} \hbox{\tencal D}\,\varphi\
\dot{\varphi}(\tau)\dot{\varphi}(0)\exp[-S_{AES}].
\end{split}
\end{gather}
Here, the subscript $W$ at the integral sign means that functional
integration is performed over phase configurations obeying the
boundary condition~\eqref{BoundCond}. The first term $\hbox{I}$ in
Eq.~\eqref{PiInst1} represents the renormalization of the partition
function due to instantons. The second term $\hbox{II}$ is the
contribution of the instanton solutions\ $\varphi_{\pm1}$\ into the
correlation function itself. The renormalized partition function
reads~\cite{panyukov,Grabert1,beloborodov1,glazman}
\begin{gather}
   \label{partition1}
  1-\sum_{W=\pm1}\frac{{\cal Z}_W}{{\cal Z}_0}=1-\frac{g^2E_c}{\pi^2T}e^{-g/2}\ln\frac{E_c}{T}\ \cos2\pi q .
\end{gather}
The contribution $\hbox{II}$ consists of two terms:
\begin{gather}
   \label{partition2}
\begin{split}
   \hbox{II}=\sum_{W=\pm1}\langle\dot{\varphi}\dot{\varphi}\rangle^{(W)}_{\omega_n}&=\sum_{W=\pm1} \langle\dot{\varphi}_W\dot{\varphi}_W\rangle^{(W)}_{\omega_n} \\
   &+ \sum_{W=\pm1}
   \langle\delta\dot{\varphi}_W\delta\dot{\varphi}_W\rangle^{(W)}_{\omega_n} ,
   \end{split}
\end{gather}
where the first term is a correlator of classical field
configurations \eqref{inst1} averaged over zero modes\ $z_a$\ and
the second term comes from fluctuations of phase $\varphi$ around
the classical solution $\varphi_W$. As shown in Appendix B the
latter term in \eqref{partition2} cancels the correction coming from
the partition function~\eqref{partition1}. Therefore,
\begin{gather}
   -\frac{\Pi(i\omega_n)}{C^2}=\langle\dot{\varphi}\dot{\varphi}\rangle^{(0)}+
   \sum_{W=\pm1}\langle\dot{\varphi}_W\dot{\varphi}_W\rangle^{(W)}_{\omega_n} .\label{PiW1}
\end{gather}

The first term in the r.h.s of Eq.~\eqref{PiW1} has been evaluated
in Sec.~\ref{Sec:First2:PT}. As it always happens in instanton
physics~\cite{polyakov}, the derivative\ $\dot{\varphi}_W(\tau)$\
coincides with a zero mode of the fluctuation\
$\delta\varphi_W(\tau)$. It is worthwhile to mention that only zero
modes of fluctuations around instanton solution contribute to the
non-perturbative renormalization of the polarization operator. The
corresponding contribution is as follows (see Appendix C for
details):
\begin{gather}
\label{phi-phi-f}
\begin{split}
    \sum_{W=\pm1}\langle\dot{\varphi}_W\dot{\varphi}_W\rangle^{(W)}_{\omega_n} &=
   4 g^2E_c \Big(\ln\frac{E_c}{T}-\frac{\pi |\omega_n|}{12 T}\Big) \\
   &\times  e^{-g/2} \cos2\pi q +\mathcal{O}(\omega_n^2) .
   \end{split}
\end{gather}
From Eqs~\eqref{PT-Pi} and \eqref{phi-phi-f} we obtain
\begin{eqnarray}
   \frac{\Pi(i\omega_n)}{C^2} &=& -\frac{2 g^2}{C} e^{-g/2} \ln \frac{E_c}{T} \ \cos 2\pi q \notag \\
   &-& 2\pi |\omega_n|\Bigl (\frac{1}{g(T)} -
   D g e^{-g(T)/2} \cos 2\pi q\Bigr )  \notag \\
   &+& \mathcal{O}(\omega_n^2), \label{PiInstF1}
\end{eqnarray}
where constant $D = (\pi^2/3) \exp(-\gamma-1)$.

%%
%%
%\subsection{\label{Sec:First2:PO}}
%%
%%

The average charge on the island can be expressed via the partition
function as
\begin{equation}\label{QdefG}
Q = q+ \frac{T}{2E_c} \frac{\partial \ln {\cal Z}}{\partial q} .
\end{equation}
Using\ Eq.~\eqref{partition1} we find the following temperature and
gate voltage dependence of the average charge in the one-instanton
approximation:
\begin{gather}\label{Qinst}
  Q=q-\frac{g^2}{\pi}e^{-g/2}\ln\frac{E_c}{T}\sin2\pi q .
\end{gather}

Performing standard analytic continuation in Eq.~\eqref{PiInstF1},
we obtain the retarded polarization operator $\Pi^R(\omega)$ in the
form of Eq.~\eqref{polar-3} with $\pi_0(T)$ satisfying
Eq.~\eqref{SWId} and
\begin{equation}
\pi_1(T) = 2\pi C^2 \Bigl (\frac{1}{g(T)} -
   D g e^{-g(T)/2} \cos 2\pi q\Bigr ) . \label{pi1one}
\end{equation}
Finally, the average energy dissipation rate will be given by
Eq.~\eqref{dissgen} with function
\begin{gather}
  \label{disone}
 \mathcal{A}(T)=\frac{2\pi C_g^2}{g(T)}
  \Bigl (1-Dg^2(T)e^{-g(T)/2}\cos2\pi q\Bigr ) .
\end{gather}
In deriving this result we changed $g$ to $g(T)$ in the factor in
front of the exponent in the r.h.s. of Eq.~\eqref{pi1one}. It is
allowed by the accuracy we are working within. Result~\eqref{disone}
asks for an interpretation. As expected, Coulomb blockade manifests
itself as a periodic dependance of dissipation\ $A(T)$\ on gate
charge\ $q$. If we ascribe this dependance to the quantum resistance
only, i.e., we write\ $\mathcal{A}(T) =C_g^2 R_q(T)$\ with\
$R_q(T)$\ following from Eq.~\eqref{disone} we face a paradox. It is
believed that  Coulomb blockade should suppress the tunneling of
electrons between the island and the lead stronger for integer
values of $q$\ than for the half-integer ones. Therefore, it would
be natural to expect that $R_q(T)$ is smaller at a half-integer
value of $q$ than at an integer one. The discussion above suggests
that we have to conceive some kind of temperature renormalization of
the gate capacitance\ $C_g$. %%
\subsection{Physical observables and gate capacitance renormalization\label{Sec:First2:PhysObser}}

As shown in Ref.~[\onlinecite{burmistrov1}], the proper physical
observables for the Coulomb blockade problem are
\begin{gather}
  \label{quantities}
  \begin{split}
    g^\prime(T)&=4\pi\Im\frac{\partial K^R(\omega)}{\partial\omega}\Big|_{\omega=0} ,\\
    q^\prime(T)&=Q+\Re\frac{\partial  K^R(\omega)}{\partial\omega}\Big|_{\omega=0},
  \end{split}
\end{gather}
where the average charge $Q$ is given by Eq.~\eqref{QdefG}, the
retarded correlation function $K^R(\omega)$ is obtained from the
Matsubara correlator
\begin{gather}
  \label{correlator}
  K(\tau_{12})=-\frac{g}{4}\alpha(\tau_{12})\big\langle
  e^{i[\varphi(\tau_1)-\varphi(\tau_2)]}\big\rangle \ \
\end{gather}
by standard analytic continuation. The physical observables
$g^\prime(T)$ and $q^\prime(T)$ describe a response of the system to
a change in the boundary condition \eqref{BoundCond}. One-instanton
contribution to the physical observables\ $g^\prime$\ and\
$q^\prime$\ reveals their periodic dependence on the external
charge\ $q$ as~\cite{burmistrov1,burmistrov2}
\begin{gather}
  \label{quantities2}
  \begin{split}
   g^\prime(T)&=g(T)\Big(1-Dg(T)e^{-g(T)/2}\cos2\pi q\Big), \\
   q^\prime(T)&=q-\frac{D}{4\pi}g^2(T)e^{-g(T)/2}\sin2\pi q .
  \end{split}
\end{gather}

Some remarks on the physical meaning of these quantities are in
order here. In the perturbative regime\ $g^\prime(T)$\ coincides
with the renormalized coupling constant\ $g(T)$\ while\
$q^\prime(T)$\ doesn't undergo any renormalization and coincides
with the external charge:\ $q^\prime(T)=q$. Thus, roughly speaking,
we can think of them as the physical observables corresponding to
the action parameters $g$ and $q$. The physics behind
quantities~\eqref{quantities} becomes even more pronounced if we
turn from a SEB to a single electron transistor (see
Fig.~\ref{figure2}). In the absence of DC-voltage between left and
right leads, a SET is described by the very same AES action
\eqref{action0}-\eqref{action3} in which the bare coupling constant\
$g=g_l+g_r$. Here, $g_{l/r}$ denotes the dimensionless conductances
of the left/right tunneling junction. The quantity\ $g^\prime(T)$\
then coincides  with the SET conductance~\cite{glazman,schon1} up to
a temperature independent factor:
\begin{gather}
  G(T)=\frac{e^2}{h} \frac{g_lg_r}{(g_l+g_r)^2}g^\prime(T).
\end{gather}
Expression for \ $q^\prime(T)$\ is possible to write in terms of
anti-symmetrized electron current-current
correlator:~\cite{burmistrov1,burmistrov2}
\begin{gather}
    \label{qPrimeDefSET}
   q^\prime(T)=Q-i\frac{(g_l+g_r)^2}{2g_lg_r}\frac{\partial}{\partial
   V_{dc}}\int_{-\infty}^0\langle[\hat I(0),\hat I(t)]\rangle\Big|_{V_{dc}=0},
\end{gather}
where $V_{dc}$ denotes the DC-voltage between the left and the right
leads and $\hat I(t) = d\hat n_d(t)/dt$\ - the current operator for
the SET.

For reasons to be explained shortly, it is natural to define the
renormalized gate capacitance
\begin{equation}
C_g(T) = \frac{\partial q^\prime(T)}{\partial U_0}. \label{CgTdef}
\end{equation}
According to Eq.~\eqref{quantities}, the quantity $C_g(T)$ is
different from the effective capacitance $\partial Q/\partial U_0$
which has been considered in the literature so far. On the
perturbative level $C_g(T)$ coincides with $C_g$; only instanton
effects make it temperature and gate voltage dependent:
\begin{gather}
    \label{eighteen}
      C_g(T)=C_g\Big(1-\frac{D}{2}g^2(T)e^{-g(T)/2}\cos2\pi q\Big) .
\end{gather}
Now, as usual, we plug in the bare capacitance\ $C_g$\ expressed
via\ $C_g(T)$\ into Eq.~\eqref{disone}. We see that instanton
corrections cancel each other and the result~\eqref{disone} for the
function $\mathcal{A}$ which determines the energy dissipation rate
becomes
\begin{equation}
\mathcal{A}(T) = \frac{2\pi C_g^2(T)}{g(T)}.
\end{equation}
With the same level accuracy we can substitute $g^\prime(T)$ for
$g(T)$ and obtain finally the following expressions for the energy
dissipation rate and the admittance in the quasi-static regime
\begin{gather}
  \label{distwo}
  \mathcal{W}_\omega=\frac{1}{2}\omega^2 C^2_g(T)R_q(T)|U_\omega|^2,\ \ R_q(T)=\frac{h}{e^2 g^\prime(T)} , \\
  \mathcal{G}(\omega) = -i\omega \frac{\partial Q}{\partial U_0}+\frac{C}{C_g}C_g^2(T)R_q(T)\omega^2. \label{admittance_q1}
\end{gather}

Several remarks are in order here. The results~\eqref{distwo} and
\eqref{admittance_q1} are valid in the weak-coupling regime:
$g^\prime(T)\gg 1$, in which the quantities $\partial Q/\partial
U_0$, $g^\prime(T)$ and $C_g(T)$ are given by Eqs~\eqref{Qinst},
\eqref{quantities2} and \eqref{eighteen}, respectively. Relations
\eqref{distwo} and \eqref{admittance_q1} fully describe the
quasi-static dynamics of SEB. The energy dissipation rate factorizes
into the product of well-defined physical observables in complete
analogy with classical expression~\eqref{admittance0}. The
admittance behavior is different from what we were expecting to get.
Indeed, its imaginary and real components involve two different
capacitances: effective capacitance $\partial Q/\partial U_0$\ and
renormalized gate capacitance $C_g(T)$. Moreover, the temperature
independent factor $C/C_g$ survives in the real part of
$\mathcal{G}(\omega)$.

%%%%%%%%%%%%%%%%%%%%%%%%%%%%%%%%%%%%%
%%%%%%%%%%%%%%%%%%%%%%%%%%%%%%%%%%%%%%
%%%%%%%%%%%%%%%%%%%%%%%%%%%%%%%%%%%%%%
%
\section{Strong coupling regime, $g\ll1$\label{Sec:Second}}
As follows from Eqs~\eqref{quantities}, the physical observables
$g^\prime(T)$ and $q^\prime(T)$ are defined for arbitrary values of
$g$. Therefore, it is of great interest to compute energy
dissipation rate and the SEB admittance in the opposite regime, of
small dimensionless tunneling conductance\ $g\ll1$. The question we
ask is whether the results \eqref{distwo} and \eqref{admittance_q1}
with the proper\ $C_g(T)$\ and\ $R_q(T)$\ hold? We mention that the
case $g\ll 1$ is a strong coupling regime from the field-theoretical
point of view. In what follows, we compute energy dissipation rate
by means of two different approaches. The first one is a refined
field-theoretical method centered around Matveev's projective
Hamiltonian.~\cite{matveev} The second one is more straightforward
approach of rate equations on which the `orthodox theory' of Coulomb
blockade was based.~\cite{kulik} We demonstrate how these two
approaches beautifully complement each other. %%
\subsection{Preliminaries}
\begin{figure}%[t]
   \includegraphics[width=65mm]{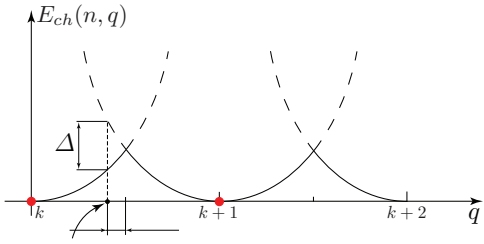}
    \caption{\label{figure3}
    (Color online) Charging energy\ $E_{ch}=E_c(n-q)^2$\ as a function of gate charge $q$.
          }
\end{figure}
%
%
%\subsubsection{Hamiltonian}

We center our effort around the most interesting case: the vicinity
of a degeneracy point:\ $q=k+1/2$\ where $k$ is an integer.
Following Ref.~[\onlinecite{matveev}], the hamiltonian
\eqref{ham1}-\eqref{ham3} can be simplified by truncating the
Hilbert space of electrons on the island to two charging states :
with\ $Q=k$\ and\ $Q=k+1$. The projected hamiltonian then takes a
form of $2\times2$\ matrix acting in the space of these two charging
states. Denoting the deviation of the external charge from the
degeneracy point by $\Delta$: $q=k+1/2-\Delta/(2E_c)$\ we write the
projected hamiltonian as:~\cite{matveev}
\begin{gather}
  \label{ham4}
   H=H_0 + H_t + \Delta S_z + \frac{\Delta^2}{4E_c}+\frac{E_c}{4}
\end{gather}
where $H_0$ is given by Eq.~\eqref{ham2} and
\begin{equation}
 H_t=\sum_{k,\alpha}t_{k\alpha} a^\dagger_k d_\alpha
   S^++\hbox{h.c.}
\end{equation}
Here, $S^z,\ S^{\pm}=S^x\pm iS^y$\ are ordinary (iso)spin\ $1/2$\
operators.
%\subsubsection{Response functions}
The presence of small AC component in the gate voltage changes the
parameter\ $\Delta$\ according to:\ $\Delta\rightarrow\Delta-(e
C_g/C)U_\omega\cos\omega t$. This time the response of the system to
AC gate voltage is determined by the isospin correlation function
$\Pi_{s}^R(\omega)$ (see Appendix A) which Matsubara counterpart is
given by
\begin{gather}
  \label{spin-correlator}
  \Pi_s(\tau)=\langle{\cal T}_\tau S^z(\tau)S^z(0)\rangle .
\end{gather}
The energy dissipation rate and SEB admittance can be expressed as
follows
\begin{gather}
\label{spin-dissipation}
\begin{split}
   \mathcal{W}_\omega&=\frac{C_g^2}{2 C^2}\omega \Im \Pi_s^R(\omega)|U_\omega|^2,\\
   \mathcal{G}(\omega)&=-i\omega\frac{C_g}{C}\Pi_{s}^{R}(\omega) .
  \end{split}
\end{gather}
Therefore, we need to proceed with the computation of\
$\Pi_{s}(\tau)$.

To deal with spin operators it is convenient to use Abrikosov's
pseudo-fermion technique.~\cite{abrikosov} We introduce
two-component pseudo-fermion operators $\psi^\dagger_\alpha$,
$\psi_\alpha$ such that
\begin{gather}
   \label{pseudofermion}
  S^i=\psi^\dagger_\alpha S^i_{\alpha\beta}\psi_\beta .
\end{gather}
Pseudo-fermions bring in the redundant unphysical states when\
$\sum_\alpha\psi^\dagger_\alpha\psi_\alpha>1$. To exclude these
states one adds an additional chemical potential $\eta$\ to the
hamiltonian. It is necessary to set\ $\eta\rightarrow-\infty$\ at
the end of any calculation. The physical partition function ${\cal
Z}$ and correlators $\langle \mathcal{O}\rangle$ can be found from
the pseudo-fermionic ones as
\begin{gather}
 \label{pf-partition}
  \begin{split}
  {\cal Z}&=\lim_{\eta\rightarrow-\infty}\frac{\partial}{\partial e^{\beta\eta}}{\cal Z}_{pf} ,\\
  \langle{\cal O}\rangle&=\lim_{\eta\rightarrow-\infty}\bigg\{ \langle{\cal O}\rangle_{pf}+
  \frac{{\cal Z}_{pf}}{{\cal Z}}\frac{\partial}{\partial e^{\beta\eta}}\langle{\cal O}\rangle_{pf}\bigg\} .
  \end{split}
\end{gather}
The elegance of pseudo-fermion technique lies in the fact that
diagrams with pseudo-fermion loops vanish when one sets:\
$\eta\rightarrow-\infty$.

Next, we plug representation~\eqref{pseudofermion} into the
hamiltonian \eqref{ham4}, switch to Matsubara basis and integrate
out electrons in the lead and the island. Done in the parametric
regime~\eqref{aes-condition2} this leads to the following effective
action:
\begin{equation}
 \label{pf-action}
  \begin{split}
   S&=
  \int_0^\beta d\tau\bar{\psi}\Big(\partial_\tau+\frac{\sigma_z \Delta}{2}-\eta\Big)\psi \\
  &+\frac{g}{4}\int_{0}^\beta d\tau_1d\tau_2\alpha(\tau_{12})[\bar{\psi}(\tau_1)\sigma_-\psi(\tau_1)]
   [\bar{\psi}(\tau_2)\sigma_+\psi(\tau_2)] \\
   &+\frac{\beta \Delta^2}{4E_c}+\frac{\beta E_c}{4} .
  \end{split}
\end{equation}
Here,\ $\sigma_i$ stand for Pauli matrices and\
$\sigma_{\pm}=(\sigma_x\pm i\sigma_y)/2$. Action similar to
Eq.~\eqref{pf-action} has been first analyzed by Larkin and Melnikov
in Ref.[\onlinecite{larkin}]. In modern terminology,
Eq.~\eqref{pf-action} corresponds to the XY case of the Bose-Kondo
model for the spin $1/2$.~\cite{Sachdev,Si,Demler} Effective
action~\eqref{pf-action} is very suitable for our purpose since it's
coupling constant\ $g\ll1$\ justifying perturbative expansion.

\begin{figure}%[t]
  \includegraphics[width=75mm]{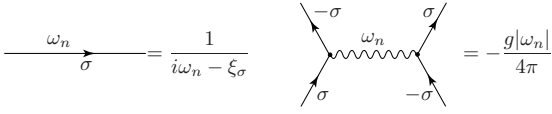}
   \caption{\label{figure4}
    Feynman rules for pseudo-fermion action;\break
    \centerline{$\xi_\sigma=-\eta+\frac{\sigma\Delta}{2}$}.
          }
\end{figure}

First, we establish the relation between pseudo-fermion and physical
partition function. From Eq.~\eqref{pf-partition}, we find
\begin{gather}
  \label{pf-partition1}
 {\cal Z} =\lim\limits_{\eta\to 0}{\cal Z}_{pf}e^{-\beta\eta}\sum_\sigma G_\sigma(\tau)\big|_{\tau\rightarrow0^-}
\end{gather}
Here, we denoted\ $G_\sigma(\tau) = -\langle {\cal T}_\tau
\psi_\sigma(\tau)\bar{\psi}_\sigma(0)\rangle$\ the exact
pseudo-fermion Green's function. The Feynman rules for action
\eqref{pf-action} are shown in Fig.~\ref{figure4}. In the zeroth
order in $g$, we obtain
\begin{equation} \label{ZerothOrderg}
G_\sigma(i\varepsilon_n) =\frac{1}{i\varepsilon_n -\xi_\sigma},
\quad {\cal Z}=2\cosh\frac{\beta\Delta}{2} , \quad {\cal Z}_{pf}=1,
\end{equation}
where $\varepsilon_n = \pi T (2n+1)$ and $\xi_\sigma=-\eta +\sigma
\Delta/2$. Spin-spin correlation function\ \eqref{spin-correlator}
written in terms of pseudo-fermions becomes
\begin{gather}
  \label{polarization1}
  \Pi_{s,pf}(\tau)=\frac{1}{4}\langle{\cal T}_\tau[\bar{\psi}(\tau)\sigma^z\psi(\tau)][\bar{\psi}(0)\sigma^z\psi(0)]\rangle ,
\end{gather}
where the average is taken with respect to action~\eqref{pf-action}.
The physical correlation function is obtained from
$\Pi_{s,pf}(i\omega_n)$ according to Eq.~\eqref{pf-partition}.

\subsection{Spin-spin correlaion function $\Pi^R_{s,pf}(\omega)$. First order in $g$.}
\begin{figure}%[t]
  \includegraphics[width=85mm]{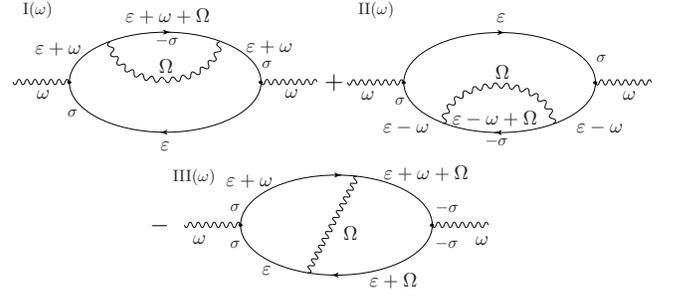}
   \caption{\label{figure5}
    Feynman diagrams defining the polarization operator in the lowest
    order.
          }
\end{figure}

We start by calculating the polarization operator
(\ref{polarization1}) in the lowest possible order of perturbation
theory. It happened that the first non-trivial contribution to\
$\Pi_{s,pf}(i\omega_n)$\ came from the first order perturbation
theory. The relevant Feynman diagrams are depicted in
Fig.~\ref{figure5}.
The computation of\ $\Pi_{s,pf}(i\omega_n)$\ is rather
straightforward and is presented in the appendix. The result is
\begin{equation}
  \label{polar-1}
  \Pi_{s,pf}(i\omega_n) = \frac{g}{4\pi^2} \frac{F^R(i\omega_n)+F^R(-i\omega_n)}{(i\omega_n)^2} e^{\beta\eta} \sinh\frac{\beta\Delta}{2}, \end{equation}
where\ $F^R(\omega)$\ is a regular in the upper half-plane of
$\omega$ function:
\begin{equation}
   F^R(\omega)=\sum_{\sigma=\pm1} \bigg[(\Delta+\sigma \omega)\psi\Big(\frac{\omega+\sigma\Delta}{2\pi T i}\Big)-
      \Delta\psi\Big(\frac{i\sigma\Delta}{2\pi T}\Big)\bigg] . \label{FR_fun}
\end{equation}
Here, $\psi(x)$\ denotes the Euler digamma-function. The analytical
continuation should be made with some care. We want to recover the
retarded polarization operator\ $\Pi_s^R(\omega)$\ which is regular
in the upper half-plane of\ $\omega$. Since\ $\psi(x)$\ has poles
at\ $x_n=-n$\ with natural\ $n$, the operator \eqref{polar-1} has
poles in both halves of a complex plane. We get rid of superfluous
ones with the help of identity:
\begin{gather}
   \psi(z)-\psi(1-z)=-\pi\cot\pi z .
\end{gather}
Using a well-known relation for real $x$
\begin{gather}
   \Im \psi(ix)=\frac{1}{2x}+\frac{\pi}{2}\coth\pi x
\end{gather}
together with Eqs~\eqref{pf-partition} and \eqref{ZerothOrderg}, we
arrive at the following expression for the imaginary part of the
polarization operator:
\begin{gather}
  \label{polar-2}
        \Im\Pi_s^R(\omega) = \frac{g}{8\pi} \bigg\{
          \frac{\Delta}{\omega^2}\sum_{\sigma=\pm1}\sigma\coth\frac{\Delta-\sigma\omega}{2T} \\
          +\frac{1}{\omega}\bigg[2\coth\frac{\Delta}{2T}-\sum_{\sigma=\pm1}\coth\frac{\Delta-\sigma\omega}{2T}\bigg]
      \bigg\} \tanh\frac{\Delta}{2T} . \notag
\end{gather}
Expression \eqref{polar-2} has a striking feature. It's divergent in
the limit\ $\omega\rightarrow0$. Indeed,
\begin{gather}
   \label{polar-div}
    \Im\Pi_s^R(\omega)=\frac{g}{4\pi\omega}\frac{\beta\Delta}{\sinh\beta\Delta},\
    \ \omega\rightarrow0 .
\end{gather}
The explanation is as follows. In essence the correlator
\eqref{spin-correlator} describes the noise of a fluctuating charge
inside a metallic island. It was computed firstly in
Ref.[\onlinecite{schon}]. The author of Ref.[\onlinecite{schon}]
however obtained a different (regular at\ $\omega\rightarrow0$)\
expression. He used a special type of analytical continuation which
yields a symmetric noise\ $\langle\{\hat n_d(t),\hat
n_d(0)\}\rangle$. We, on the other hand, are interested in its
antisymmetric counterpart which is the response function\
\eqref{polar-0}. It is exactly this retarded antisymmetric function
which is obtained via standard analytical continuation procedure.

The unphysical divergency~\eqref{polar-div} comes from the
non-trivial and essentially non-perturbative infrared structure of a
polarization operator\ $\Pi_s^R(\omega)$. In what follows we prove
that the partial summation of some infinite classes of diagrams
resolves this singularity yielding the result:
\begin{gather}
  \label{structure}
   \Im \Pi_s^R(\omega)\sim\frac{g\omega}{z^2+\omega^2} ,
   %\rightarrow \frac{\omega}{g\Delta^2},\ \omega\rightarrow0
\end{gather}
where $z \sim g \Delta$ at $T=0$. As seen from Eq.~\eqref{structure}
the limits\ $\omega\rightarrow0$ and $g\rightarrow0$\ do not commute
which explains how the artificial divergency in
Eq.~\eqref{polar-div} arises. Now we proceed with a more accurate
computation of the correlator\ $\Pi_s^R(\omega)$.

\subsection{One-loop structure of the pseudo-fermion theory}
Throughout all our computation we will need some knowledge of the
one-loop logarithmic structure of the pseudo-fermion theory.
%One loop correction to Green's functions was computed in [\onlinecite{burmistrov2}].
The bare Green's function is modified by the self-energy:
\begin{equation}
G_{\sigma}(i\varepsilon_n)=\frac{1}{i\varepsilon_n-\xi_\sigma-\Sigma_\sigma(i\varepsilon_n)}
.
\end{equation}
The leading logarithmic approximation corresponds to one-loop
renormalization. As it is known~\cite{larkin}, one extracts
self-energy \ $\Sigma_\sigma(i\varepsilon_n)$\ from a
self-consistent Dyson equation:
\begin{gather}
    \Sigma_\sigma(i\varepsilon_n)=-\frac{g}{4\pi}T\sum_{\omega_m} |\omega_m| G_{-\sigma}(i\varepsilon_n+i\omega_m) .
\end{gather}
Here, we introduce $\omega_m=2\pi T m$. The vital observation
[\onlinecite{larkin,Si,Demler}] is that the action (\ref{pf-action})
can be renormalized with only one scaling parameter\ $Z$. Performing
standard analytic continuation, we find~\cite{larkin,burmistrov2}
\begin{gather}
  \label{green1_0}
   G_\sigma^{R,A}(\varepsilon)=\frac{Z(\lambda)}{\varepsilon-\bar{\xi}_\sigma\pm i\bar{g}\Gamma_\sigma(\varepsilon)},\\
   Z(\lambda)=\Bigl (1+\frac{g}{2\pi^2}\lambda \Bigr )^{-1/2}, \quad \lambda= \ln\frac{E_c}{\hbox{max}\{T,|\bar{\Delta}|,|\varepsilon|\}} . \notag
\end{gather}
Here, $\bar{\xi}_\sigma=-\eta+\sigma\bar{\Delta}/2$. $\bar{g}=g
Z^2(\lambda)$ and $\bar{\Delta}=\Delta Z^2(\lambda)$ are the
renormalized coupling constant and gap, respectively. As we see the
energy\ $E_c$\ plays the role of a reference energy scale. The
important feature of Green's function \eqref{green1_0} is its
acquired width
\begin{gather}
  \label{width}
  \Gamma_\sigma(\varepsilon)=
   \frac{1}{8\pi}(\varepsilon-\bar\xi_{-\sigma})\frac{\cosh\frac{\varepsilon}{2T}}{\sinh\frac{\varepsilon-\bar\xi_{-\sigma}}{2T}\cosh\frac{\bar\xi_{-\sigma}}{2T}}.
\end{gather}
According to the hierarchy of energy scales considered in the paper,
$E_c\gg T$ and, therefore, the logarithmic corrections\ $\sim g\ln
E_c/T$\ are not small and require special care. To get read of large
logarithms we change the reference scale of the field
theory~\eqref{pf-action} from\ $E_c$\ to\ $T$. With the help of
result \eqref{green1_0} we may rewrite the theory in terms of
renormalized fields and running coupling constants:\
$\psi_\sigma\rightarrow\sqrt{Z(\lambda)}\psi^r_\sigma$,
$g\rightarrow\bar{g}$ and $\Delta\rightarrow\bar{\Delta}$. The
action then becomes
\begin{gather}
 \label{pf-action-r}
    S[\bar{\psi},\psi,\Delta,g]=
    S[\bar{\psi}^r,\psi^r,\bar\Delta,\bar{g}]+\delta S_{c.t.},
\end{gather}
where\ $\delta S_{c.t.}$\ stands for the action of counter terms. It
is responsible for a consistent regularization of higher order (in\
$\bar g$) corrections to the physical observables. Action
(\ref{pf-action-r}) is very suitable for our purpose. All large
logarithms are absorbed  into coupling constants and fermionic
fields. This allows us to drop counter-terms in what follows. To
bind the observables defined at the reference scale\ $E_c$\ with the
renormalized ones we shall need to establish scaling of the
pseudo-fermion density\
$\rho_{pf}=\sum_\sigma\langle\bar{\psi}_\sigma\psi_\sigma\rangle$\
and $z$-component of the total spin density
$s^z_{pf}=(1/2)\sum_\sigma
\sigma\langle\bar{\psi}_\sigma\psi_\sigma\rangle$. As follows from\
Ref.~[\onlinecite{larkin}], the pseudo-fermion density\ $\rho$\ has
no scaling dimension of its own:
\begin{gather}\label{density-scaling}
\rho_{pf} =
\sum_\sigma\langle\bar{\psi}^r_\sigma\psi^r_\sigma\rangle
%  \rho(\Delta)+\delta\rho(g,\Delta)=\rho(\bar\Delta)
\end{gather}
where now the average is taken with respect to
action~\eqref{pf-action-r}. The total spin density\ $s_z$\ has the
same structure as the term proportional to \ $\Delta$\ in action
\eqref{pf-action}. Therefore, it must have the same scaling
dimension:
\begin{gather}
  \label{spin-scaling}
s^z_{pf} = \frac{1}{2} Z^2(\lambda)\sum_\sigma \sigma
\langle\bar{\psi}^r_\sigma\psi^r_\sigma\rangle,
  %   s_z(\Delta)+\delta s_{z}(g,\Delta)=Z^2(\lambda)s_z(\bar{\Delta})
\end{gather}
where again the average is taken with respect to the
action~\eqref{pf-action-r}. For completeness we present the rigorous
proof  of Eq.~\eqref{spin-scaling} via Callan-Symanzik equation in
Appendix D.

\subsection{Dyson equation for the spin-spin correlation function\ $\Pi^R_{s,pf}(\omega)$.}
\begin{figure}[t]
  \includegraphics[width=70mm]{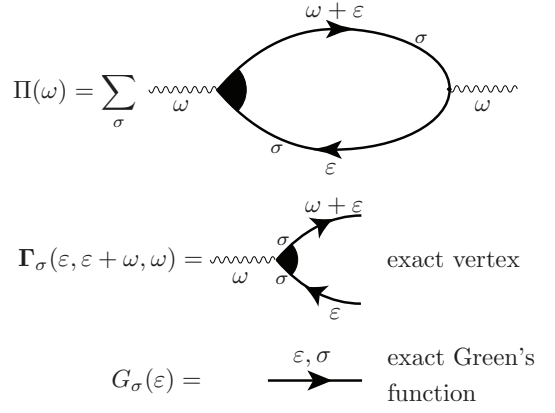}
   \caption{\label{figure6}
    Dyson equation for polarization operator\
    $\Pi_{s,pf}(i\omega_n)$.
          }
\end{figure}
The graphical representation of Dyson equation for the spin-spin
correlation function
\begin{eqnarray}
\Pi_{s,pf}(i\omega_n)&=&\frac{T}{4}\sum_{\varepsilon_k,\sigma}\bm{\Gamma}_{\sigma}(i\varepsilon_k,i\varepsilon_k+i\omega_n,i\omega_n)\\
&\times &G_{\sigma}(i\varepsilon_k)
G_{\sigma}(i\varepsilon_k+i\omega_n)  \notag
\end{eqnarray}
is illustrated in Fig.~\ref{figure6}. Here,
$\bm{\Gamma}_{\sigma}(i\varepsilon_k,i\varepsilon_k+i\omega_n,i\omega_n)$
denotes the exact vertex function.
Performing the analytic continuation in the spirit of
Ref.~[\onlinecite{eliashberg}], we find (see Appendix D for details)
\begin{widetext}
 \begin{gather}
    \label{polarization0}
   \Pi^{R}_{s,pf}(\omega)=- \sum_\sigma\int\frac{d\varepsilon}{16\pi i}
   \Big\{\bm{\Gamma}^{ARR}_\sigma(\varepsilon,\varepsilon+\omega,\omega)G^{A}_{\sigma}(\varepsilon) G^{R}_{\sigma}(\varepsilon+\omega)
   \Big[\tanh\frac{\varepsilon+\omega}{2T}-\tanh\frac{\varepsilon}{2T}\Big] \\
+
\bm{\Gamma}^{RRR}_\sigma(\varepsilon,\varepsilon+\omega,\omega)G^R_\sigma(\varepsilon)G^R_\sigma(\varepsilon+\omega)
    \tanh\frac{\varepsilon}{2T}
   -\bm{\Gamma}^{AAR}_\sigma(\varepsilon,\varepsilon+\omega,\omega)G^A_\sigma(\varepsilon) G^A_\sigma(\varepsilon+\omega)\tanh\frac{\varepsilon+\omega}{2T} \notag
   \Big\} .
 \end{gather}
 \end{widetext}
The most important task is to single out singular at\
$\omega\rightarrow0$ and $\bar{g}\to 0$ \ terms in
\eqref{polarization0}. We shall treat them separately to avoid
divergencies. Firstly, we recall that we are interested in the
quasi-static limit. Therefore, we shall proceed under assumptions
$\omega \ll \max\{\bar{\Delta},T\}$ but $\omega \sim
\bar{g}\max\{\bar{\Delta},T\}$. It is intuitively clear that a
singular contribution always comes from the first term in the r.h.s.
of Eq.~\eqref{polarization0} which involves the product\
$G_\sigma^AG_\sigma^R$. Indeed, we observe that the pole structure
of\ $G_\sigma^AG_\sigma^R$\ always leads to a singular denominator
of the type $(\omega+2 i\bar{g}\Gamma_\sigma)$\ as a result of
integration. This happens due to the proximity of poles in \
$G_\sigma^R$\ and\ $G_\sigma^A$. In contrast, the other terms with \
$G_\sigma^RG_\sigma^R$\ and\ $G_\sigma^AG_\sigma^A$\ are regular at\
$\bar{g}=\omega=0$\ and, therefore, free of divergencies. We may
compute their contribution setting\ $\bar{g}=0$\ and safely
expanding the result in\ $\omega$. The integrand in
Eq.~\eqref{polarization0} also has a series of Matsubara-type poles
due to the presence of hyperbolic functions. These poles lead to
logarithmically  divergent sums. The latter are controlled by the
renormalization scheme. In our case all leading logarithms are
absent. They have already been absorbed into renormalized constants\
$\bar{g}$,\ and  $\bar{\Delta}$ by the proper choice of reference
energy scale. Thus we can drop all divergent sums over Matsubara
frequencies.

%Now we compute separately integrals over regular and singular parts.
Perfoming integration over $\varepsilon$ in
Eq.~\eqref{polarization0} and expanding in\ $\omega$ where it is
allowed, we are able to write down a much simpler expression for\
$\Pi_{s,pf}^R(\omega)$:
\begin{equation}
 \label{pseudofermion2}
   \Pi_{s,pf}^{R}(\omega)=\sum_\sigma\frac{\beta}{16\cosh^2\frac{\bar\xi_\sigma}{2T}}\bigg\{1-
   \frac{\omega\bm{\Gamma}^{ARR}_\sigma(\bar\xi_\sigma,\bar\xi_\sigma+\omega,\omega)}{\omega+2i\bar{g}\Gamma_\sigma(\bar\xi_\sigma)}\bigg\} .
\end{equation}

\begin{figure}%[t]
  \includegraphics[width=85mm]{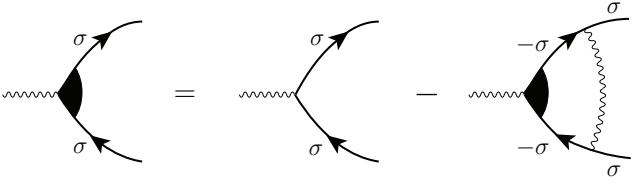}
   \caption{\label{figure7}
   Dyson equation for the vertex\ $\bm{\Gamma_\sigma}$.
          }
\end{figure}

Now we need to compute the vertex function\
$\bm{\Gamma}_\sigma^{ARR}(\varepsilon,\varepsilon+\omega,\omega)$.
The vertex function
$\bm{\Gamma}_\sigma(i\varepsilon_k,i\varepsilon_k+i\omega_n,i\omega_n)$
satisfies Dyson equation
\begin{widetext}
\begin{gather}
  \label{vertex1}
       \bm{\Gamma}_\sigma(i\varepsilon_k,i\varepsilon_k+i\omega_n,i\omega_n)=1+\frac{\bar{g} T}{4\pi}\sum_{\omega_m} |\omega_m| G_{-\sigma}(i\varepsilon_k+i\omega_m)
% \notag   \\ \times
G_{-\sigma}(i\varepsilon_k+i\omega_m+i\omega_n)
       \bm{\Gamma}_{-\sigma}(i\varepsilon_k+i\omega_m,i\varepsilon_k+i\omega_m+i\omega_n,i\omega_n)
\end{gather}
\end{widetext}
which is shown in Fig.~\ref{figure7}.
The details of analytical continuation are described in Appendix D
where we prove that the Dyson equation for the vertex
$\bm{\Gamma}_\sigma^{ARR}(\varepsilon,\varepsilon+\omega,\omega)$
becomes
\begin{gather}
   \bm{\Gamma}^{ARR}_{\sigma}(\varepsilon,\varepsilon+\omega,\omega)=1-\frac{\bar{g}}{8\pi}\int\frac{dx}{2\pi} \bm{\Gamma}_{-\sigma}^{ARR}(x,x+\omega,\omega)\notag \\
  \times G^A_{-\sigma}(x)G^R_{-\sigma}(x+\omega) (x-\varepsilon)
 \Big[2 \coth\frac{x-\varepsilon}{2T}\notag \\
 -\tanh\frac{x+\omega}{2T}-
 \tanh\frac{x}{2T}\Big] . \label{Dyson1_0}
\end{gather}
To solve it we have to make some self-consistent guess. The apparent
difficulty is that apart from singular factor
$G_\sigma^RG_\sigma^A$\ the integrand in the r.h.s. of
Eq.~\eqref{Dyson1_0} may have unknown poles coming from the vertex
function\ $\bm{\Gamma}_{-\sigma}^{ARR}$. We conjecture that these
poles result in the contribution of order of unity and are small
comparing to singular contribution from the product\
$G_\sigma^RG_\sigma^A$. Hence we may perform an integral in the
r.h.s. of Eq.~\eqref{Dyson1_0} and arrive at the following result
\begin{gather}
   \bm{\Gamma}^{ARR}_{\sigma}(\varepsilon,\varepsilon+\omega,\omega)=1-\frac{\bar{g}}{8\pi} \frac{\bar\xi_{-\sigma}-\varepsilon}{\sinh\frac{\bar\xi_{-\sigma}-\varepsilon}{2T}} \Biggl \{\frac{\cosh\frac{\varepsilon}{2T}}{\cosh\frac{\bar\xi_{-\sigma}}{2T}}
 \notag \\
  + \frac{\cosh\frac{\varepsilon+\omega}{2T}}{\cosh\frac{\bar\xi_{-\sigma}+\omega}{2T}}\Biggr \}
   \frac{\bm{\Gamma}^{ARR}_{-\sigma}(\bar\xi_{-\sigma},\bar\xi_{-\sigma}+\omega,\omega)}{-i\omega+2 \bar{g}\Gamma_{-\sigma}(\bar\xi_{-\sigma})}  . \label{vertex-solution}
\end{gather}
We see that the solution does have an additional series of poles in
the $\varepsilon$ plane. However these are Matsubara-type poles and
are irrelevant as was argued above. Setting the external energy\
$\varepsilon=\bar\xi_\sigma$\ we obtain the self-consistent equation on\
$\bm{\Gamma}^{ARR}_{\sigma}(\bar\xi_\sigma,\bar\xi_\sigma+\omega,\omega)$.\
%and proceed to solve it in the leading order in\ $g$\ and small frequencies\ $\om\ll g\{T,\Delta\}$.
The solution reads
\begin{gather}
  \label{vertex-solution1}
\bm{\Gamma}^{ARR}_\sigma(\bar\xi_\sigma,\bar\xi_\sigma+\omega,\omega)=\frac{1}{\omega}
   \frac{\big[\omega+2i\bar{g}(\Gamma_{-\sigma}-\Gamma_{\sigma})\big]\big[\omega+2i\bar{g}\Gamma_\sigma\big]}{\omega+2i\bar{g}(\Gamma_{-\sigma}+\Gamma_{\sigma})}
\end{gather}
Here,\ $\Gamma_\sigma=\Gamma_\sigma(\bar\xi_\sigma)$\ is the width of
the Green's function defined in Eq.~\eqref{width}.

Collecting Eqs~\eqref{pseudofermion2} and \eqref{vertex-solution} we
write down the result for the spin-spin correlation function
\begin{gather}
  \label{polarization2_0}
     \Pi_{s,pf}^R(\omega)=\frac{\bar{g}}{4\pi} \frac{\bar{\Delta}}{T\sinh\frac{\bar{\Delta}}{2T}} \left [-i\omega+\frac{\bar{g}\bar{\Delta}}{2\pi}\coth\frac{\bar{\Delta}}{2T} \right ]^{-1} .
\end{gather}
Finally, taking into account Eqs~\eqref{pf-partition1} and
\eqref{ZerothOrderg}, we obtain the following result for the
spin-spin correlator:
\begin{gather}
  \label{polarization2}
     \Pi_s^R(\omega)=\frac{\bar{g}Z^4}{4\pi}\frac{\bar{\Delta}}{T\sinh\frac{\bar{\Delta}}{T}}\left [ -i\omega+\frac{\bar{g}\bar{\Delta}}{2\pi}\coth\frac{\bar{\Delta}}{2T}\right ]^{-1} .
\end{gather}
Here we restored factor\ $Z^4$\ which provides a correct scaling
dimension of spin fields according to Eq.~\eqref{spin-scaling}.
\subsection{The admittance and the energy dissipation rate}

With the help of Eq.~\eqref{spin-dissipation} we obtain the
admittance of the SEB for frequencies\ $\omega \ll
\max\{\bar{\Delta},T\}$:
\begin{gather}
  \label{admittance_strong}
  \mathcal{G}(\omega)=\frac{C_g}{C}\frac{\bar{g}Z^4}{4\pi}\frac{\bar{\Delta}}{T\sinh\frac{\bar{\Delta}}{T}}\frac{-i\omega}{ -i\omega+\frac{\bar{g}\bar{\Delta}}{2\pi}\coth\frac{\bar{\Delta}}{2T}} .
 \end{gather}
The average charge $Q$ and the physical observables $g^\prime(T)$
and $q^\prime(T)$ can be found from the pseudo-fermion
theory~\eqref{pf-action} if one substitutes the transverse spin-spin
correlation function
 \begin{equation}
K_s(\tau_{12}) =-\frac{g}{4}\alpha(\tau_{12})\langle
S^+(\tau_1)S^-(\tau_2)\rangle
\end{equation}
for $K(\tau_{12})$ in
Eq.~\eqref{quantities}.~\cite{burmistrov1,burmistrov2} In the
leading logarithmic approximation, the average charge and the
physical observable $g^\prime$ are given by~\cite{schoeller}
\begin{gather}
Q(T)=k+\frac{1}{2}-\frac{Z^2}{2}\tanh\frac{\bar{\Delta}}{2T} ,\label{Qpf}\\
g^\prime(T) =
\frac{\bar{g}}{2}\frac{\bar{\Delta}}{T\sinh\frac{\bar{\Delta}}{T}}.\label{gPpf}
\end{gather}
The temperature dependence of the other physical observable
$q^\prime$ is as follows~\cite{burmistrov2}
\begin{gather}
  q^\prime(T)=k+\frac{1}{2}-\frac{1}{2}\tanh\frac{\bar{\Delta}}{2T} .\label{qPpf}
\end{gather}

To get the energy dissipation rate we expand
expression~\eqref{admittance_strong} in  $\omega$. Using the
identity $d\bar{\Delta} = - Z^2 dU_0/C$ and
Eqs~\eqref{Qpf}-\eqref{qPpf}, we obtain the energy dissipation rate
and the admittance of the SET in the quasi-static regime:
\begin{gather}
  \label{dis_str}
  \mathcal{W}_\omega=\frac{1}{2}\omega^2 C^2_g(T)R_q(T)|U_\omega|^2,\ \ R_q(T)=\frac{h}{e^2 g^\prime(T)} , \\
  \mathcal{G}(\omega) = -i\omega \frac{\partial Q}{\partial U_0}+\frac{C}{C_g}C_g^2(T)R_q(T)\omega^2. \label{admittance_q1_str}
\end{gather}
Here, the renormalized gate capacitance and the effective
capacitance becomes
\begin{gather}
     C_g(T)=\frac{\partial q^\prime}{\partial U_0} = C_g \frac{Z^2}{2}\frac{E_c
     }{T\cosh^2\frac{\bar{\Delta}}{2T}},\label{CgT_str}\\
     \frac{\partial Q}{\partial U_0} = C_g \frac{Z^4}{2}\frac{E_c
     }{T\cosh^2\frac{\bar{\Delta}}{2T}} .\label{C_effT_str}
\end{gather}

Several remarks are in order here. The results~\eqref{dis_str} and
\eqref{admittance_q1_str} are valid in the strong-coupling regime,
$g\ll 1$ and near the degeneracy point $|\Delta|\ll E_c$. The
accuracy we are working with (the leading logarithmic approximation)
allows us to make the following key observation. The expressions for
energy dissipation rate~\eqref{dis_str} and
admittance~\eqref{admittance_q1_str} cast in terms of the
quantities\ $\partial Q/\partial U_0$,\ $g^\prime(T)$\ and\
$C_g(T)$\ coincide with the ones obtained in the weak-coupling
regime. It makes us suggest that the results~\eqref{dis_str} and
\eqref{admittance_q1_str} are valid for all temperature range
$E_c\gg T\gg \delta$ and all values of $g$.

We mention that formula~\eqref{dis_str} for $R_q(T)$ is a truly
non-perturbative in $g$\ result. Despite obvious complications we
overcame to obtain it, the expression for\ $g^\prime(T)$\ (stripped
of all logarithmic scaling) is the same as obtained in a much
simpler approach of sequential tunneling. This approach known as the
`orthodox theory' of a Coulomb blockade will help us to shed light
on the physical meaning of results~\eqref{dis_str} and
\eqref{admittance_q1_str}. Further calculations are formulated in
the language of rate equations~\cite{kulik} which lie in the basis
of the orthodox theory.
\subsection{Rate equations approach}
The rate approach is less general since it is essentially a Fermi
golden rule approximation. It overlooks virtual processes and is
unable to reproduce logarithmic scaling of physical observables. On
the other hand rate equations are much easier to solve than
corresponding Dyson equations used above in a field-theoretical
treatment. We are going to demonstrate that rate equations allow us
to find the admittance for frequencies which are not restricted by
the condition\ $\omega\ll\max\{\bar{\Delta},T\}$\ imposed by the
field approach. Eventually we will conceive a prescription on how
the admittance formula \eqref{admittance_strong} can be generalized
for arbitrary (but still not very large\ $\omega\ll E_c$)
frequencies.

As above, only two charging states of the island are counted. We
denote them as follows: state 0 corresponds to the average charge\
$Q=k$\ and state 1 corresponds to \ $Q=k+1$. Probabilities for each
state are denoted as\ $p_0$\ and\ $p_1$ which satisfy $p_0+p_1=1$.
Master equation has the standard form:~\cite{kulik}
\begin{gather}
   \label{master1}
   \dot{p}_0=-\Gamma_{10}p_0+\Gamma_{01}p_1 .
\end{gather}
Here, $\Gamma_{01}$\ and\ $\Gamma_{10}$\ are tunneling rates {\it
from} and {\it to} the metallic island, respectively. We should keep
in mind that tunneling rates\ $\Gamma_{01/10}$\ are proportional to
dimensional conductance of the tunneling junction\ $g$ which is the
expansion parameter of our problem. The average current through the
contact is
%\begin{gather}
$I=-\dot{p}_0$.
%\end{gather}
Since we are interested in the linear response of the current to the
time-dependent gate voltage $U(t)=U_0+U_\omega\cos\omega t$, we
expand the tunneling rates to the first order in amplitude\
$U_\omega$:
\begin{gather}
%  \begin{split}
   \Gamma_{01/10}(t)=\Gamma^0_{01/10}+\frac{C_gU_\omega}{2C}\Bigl [ \gamma_{01/10}(\omega) e^{-i\omega t} \\
   + \gamma_{01/10}(-\omega) e^{i\omega t}  \Bigr ]. \notag
%   \Gamma_{10}&=\G^0_{10}+U_0\frac{C_g}{C}\gamma_{10}(t)
%  \end{split}
\end{gather}
Then it is easy to find the following relation for the admittance:
\begin{gather}
    \label{current1}
    \mathcal{G}(\omega)=-i\omega\frac{C_g}{C}\frac{\gamma_{10}(\omega)\Gamma^0_{01}-\gamma_{01}(\omega)\Gamma^0_{10}}{(\Gamma^0_{01}+\Gamma^0_{10})(-i\omega+\Gamma^0_{01}+\Gamma^0_{10})} .
\end{gather}
The equilibrium tunneling rates are well-known~\cite{kulik}
\begin{gather}
   \label{gamma-big}
    \Gamma^0_{01/10}=\frac{g\Delta}{4\pi}\Big(\coth\frac{ \Delta}{2T} \pm 1\Big) .
\end{gather}
We mention that up to the logarithmic corrections\ $\Gamma^0_{01/10}
= 2 g \Gamma_\pm$. A straightforward calculation of the tunneling
rates yields (see Appendix E)
\begin{gather}
   \label{gamma-small-complete}
   \gamma_{01/10}(\omega)=\mp\frac{g}{4\pi}\bigg[1\pm \frac{i}{\pi\omega}F^R(\omega)\bigg]
\end{gather}
where function\ $F_R(\omega)$\ was introduced in equation
\eqref{FR_fun}. Plugging
Eqs~\eqref{gamma-big}-\eqref{gamma-small-complete} into
Eq.~\eqref{current1} we arrive at the general expression for
admittance:
\begin{gather}
  \label{admittance_rate0}
   \mathcal{G}(\omega)=\frac{C_g}{C}\frac{g}{4\pi\coth\frac{\beta\Delta}{2}}\frac{-i \omega\coth\frac{\Delta}{2T}-\frac{1}{\pi}F^R(\omega)}
   {-i\omega+\frac{g\Delta}{2\pi}\coth\frac{\Delta}{2T}} .
\end{gather}
In  order to relate this result to field-theoretical
result~\eqref{admittance_strong}, we expand the function
$F^R(\omega)$ to the first order in\ $\omega$:
\begin{gather}
    F^R(\omega) =i\pi \omega \left
  (\frac{\Delta}{2T\sinh^2\frac{\Delta}{2T}}-\coth\frac{\Delta}{2T}\right) + \mathcal{O}(\omega^2).
\end{gather}
Plugging this into \eqref{current1} we get the familiar expression:
\begin{gather}
  \label{admittance_rate}
  \mathcal{G}(\omega)=\frac{C_g}{C}\frac{g}{4\pi}\frac{\beta\Delta}
  {\sinh\beta\Delta}\frac{-i\omega}{-i\omega+\frac{g\Delta}{2\pi}\coth\frac{\Delta}{2T}}
\end{gather}
which is valid for $\omega \ll \max\{\Delta,T\}$ and nearly repeats
 result~\eqref{admittance_strong} for the admittance. The only
difference is in the scaling factor\ $Z$\ which is absent in the
rate equations approach. Now we may easily guess a prescription on
how to generalize Eq.~\eqref{admittance_strong} for arbitrary\
$\omega$. A correctly defined observable admittance ought to scale
as\ $Z^4$. It should also be expressed in terms of the renormalized
parameters\ $\bar{g}$\ and\ $\bar{\Delta}$\ only. This leads us to
the following result
\begin{gather}
  \label{admittance_rate10}
   \mathcal{G}(\omega)=\frac{C_g}{C}\frac{\bar{g} Z^4}{4\pi\coth\frac{\bar\Delta}{2T}}
   \frac{-i\omega\coth\frac{\bar\Delta}{2T}+\frac{1}{\pi}{\bar F}^R(\omega)}
   {-i\omega+\frac{\bar{g}\bar\Delta}{2\pi}\coth\frac{\bar\Delta}{2T}}
\end{gather}
which as we believe describes the admittance for $\omega\ll E_c$ in
the strong-coupling regime $g\ll 1$. Here,\ the function ${\bar
F}^R(\omega)$ is given by $F_R(\omega)$ in which $\bar\Delta$ is
substituted for $\Delta$. Finally, we mention that at finite
frequency\ $\omega$\ the parameter $\lambda$ which determines the
scaling parameter $Z$ in Eq.~\eqref{green1_0} should be modified as
follows:
\begin{gather}
  \lambda = \ln\frac{E_c}{\max\{T,|\bar\Delta|,|\omega|\}} .
\end{gather}

%%%%%%%%%%%%%%%%%%%%%%%%%%%%%%%%%%%%%
%%%%%%%%%%%%%%%%%%%%%%%%%%%%%%%%%%%%%%
%%%%%%%%%%%%%%%%%%%%%%%%%%%%%%%%%%%%%%
\section{Discussions and conclusions\label{Sec:Last}}

%. Идея про общность формулы C_g^2(T) R_q(T). 2. Связь q^\prime и С_g (антисимметризованный коррелятор), %обсуждение зависимости диссипации от q (физическая причина а-ля Delsing) 3. Сравнение с внутренней %диссипации (макроскопическая формула) разные режимы по частоте 4. Возможная нелинейная зависимость %(замена T на \omega) 5. Обсуждение эксперимента Delsing 6. Легкая критика Бутиккера

As we demonstrated in the previous sections the energy dissipation
rate $\mathcal{W}_\omega$ is given by Eq.~\eqref{eq1} with $R_q(T) =
h/e^2 g^\prime(T)$ and $C_g(T) = \partial q^\prime(T)/\partial U_0$
in both weak and strong coupling regimes. We emphasize that the
physical observables  $g^\prime$ and $q^\prime$ are defined in terms
of the correlation function of the phase field $\varphi(\tau)$ of
the AES model (see Eq.~\eqref{quantities}). Therefore, they can be
found, in general, not only in weak and strong coupling regimes but
for arbitrary values of $g$ and $q$. Hence, it is natural to assume
that Eq.~\eqref{eq1} as well as Eq.~\eqref{admittance_q1} hold in
general for a SEB under conditions of applicability of the AES model
which are $E_{\rm Th}\gg E_c\gg T\gg \delta\max\{1,g\}$ and $
g/N_{ch}\ll1$. Although, at present we are not able to prove this
conjecture we believe strongly it is true.

Originally,~\cite{burmistrov1} the physical quantity $q^\prime(T)$
has been introduced for a SET and its physical meaning was
interpreted in terms of the average charge on the island and the
antisymmetrized current-current correlation function (see
Eq.~\eqref{qPrimeDefSET}. If we introduce non-symmetrized current
noise in the SET~\cite{ImryBook,BlanterReview}
\begin{equation}
S_I(\omega,V_{dc}) = \int\limits_{-\infty}^\infty dt \, e^{-i\omega
t} \langle \hat{I}(t) \hat{I}(0) \rangle ,
\end{equation}
then we can present Eq.~\eqref{qPrimeDefSET} as
\begin{gather}
q^\prime = Q + \frac{(g_l+g_r)^2}{2\pi g_l g_r}  p.v. \int
\limits_{-\infty}^\infty \frac{d\omega}{\omega} \frac{\partial
S_I(\omega,V_{dc})}{\partial V_{dc}} \Biggr |_{V_{dc}=0}\label{qSI}
.
\end{gather}
Therefore,  to measure $q^\prime(T)$ two separate experiments are
needed:  the measurement of the average charge on the island at
$V_{dc}=0$ and the measurement of the non-symmetrized current noise
$S_I(\omega,V_{dc})$. Although experimental designs probing the
non-symmetrized current noise have already been
proposed,~\cite{Lesovik} and measurements have recently been taken
from a number of electronic quantum devices~\cite{Deblock}, it is
still a challenge. Our present results indicate that the quantity\
$q^\prime$\ can be related to the renormalized gate capacitance\
$C_g(T)$. Namely, $C_g(T)=\partial q^\prime/\partial U_0$, provided
that result~\eqref{eq1} holds in general (not  in a weak and strong
coupling regimes only). The latter capacitance can be extracted from
measurements of the energy dissipation rate and the SET conductance.

\begin{figure}[ht]
  \includegraphics[width=70mm]{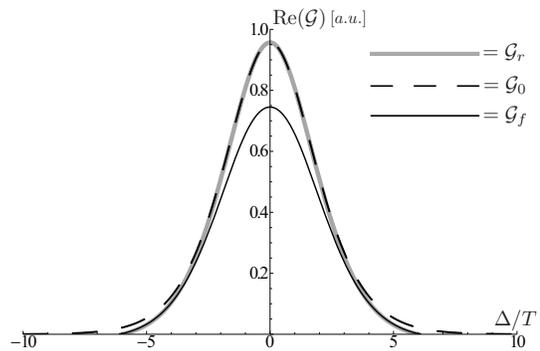}
  \caption{\label{figure8}
    The dissipative part of admittance of the SEB at fixed\ $\omega$\ as a function of\ $\Delta$.
    Three plots using three different formulae are presented.
    $G_r$\ is given by~\eqref{admittance_rate0},\ $G_0$\ is given by~\eqref{admittance_rate},
    $G_f$\ is given by~\eqref{admittance_rate10}. We use $g=0.5$, $E_c=10 T$, and $\omega = 0.8 T$.
          }
\end{figure}

Recently, the energy dissipation rate of SEB has been addressed
experimentally via radio-frequency reflectometry measurements (by
sending a continuous radio-frequency signal to the
device).~\cite{delsing} The temperature and external charge
dependences of the quantity $[\omega^2 \mathcal{A}(T)]^{-1}$ were
studied. The latter quantity was refered to as the `Sisyphus'
resistance by the authors of Ref.~[\onlinecite{delsing}]. In the
experiment the tunneling conductance was estimated to be equal
$g\approx 0.5$ such that the SEB was in the strong coupling regime.
In Ref.~[\onlinecite{delsing}] the `Sisyphus' resistance was
estimated theoretically within the rate equation approach (see
Eq.(4) in Ref.~[\onlinecite{delsing}]). Their result corresponds to
our result~\eqref{admittance_rate} for the admittance. However, our
final result for the admittance \eqref{admittance_rate10} is more
general then Eq.(4) of Ref.~[\onlinecite{delsing}]. The latter does
not take into account not only the logarithmic renormalizations of
the SEB parameters but also deviation of the function $F^R(\omega)$
from the linear one. Although the values of the SEB parameters
reported in Ref.~[\onlinecite{delsing}] are such that the difference
of the scaling factor $Z$ from unity is several per cent,
logarithmic renormalizations in the expression for the admittance
yield noticable effect. This is  shown in Fig.~\ref{figure8}. In
addition, the function $F^R(\omega)$ can be written as the linear
one only for frequencies $\omega \ll \max\{\Delta,T\}$ which is not
the case for the low temperature data of
Ref.~[\onlinecite{delsing}]. Therefore, the experimental data of
Ref.~[\onlinecite{delsing}] needs to be reanalyzed with the help of
Eq.~\eqref{admittance_rate10}.

The authors of Ref.~[\onlinecite{delsing}] claim that their results
for the `Sysiphus' resistance indicate the violation of the
Kirchhoff’s laws. They argue that the admittance they measure does
not correspond to the equivalent circuit of SEB with bare values of
the gate capacitance $C_g$ and the tunneling conductance $g$.
However, by the same logic one could claim the violation of the
Kirchhoff's laws in measurements of the SET conductance
$\mathcal{G}(T)$ because it is different from $g_lg_r/(g_l+g_r)$.
Our results imply that the energy dissipation rate (inverse of the
`Sysiphus' resistance) in the SET can be  obtained from the
Kirchhoff's laws if one substitute $C_g $ and $g$ for $C_g(T)$ and
$g^\prime(T)$ in the equivalent circuit.
%Delsing \omega\sim 30 mK
%E_c\sim 0.7 K
%%%%%%%%%%%%%%%%%%%%%%%%%%%%%
As one can see from Fig.~\ref{figure8}, the energy dissipation rate
is maximal for $\Delta=0$ which corresponds to the half-integer
values of the external charge $q$. It occurs because the larger
value of $\Delta$ is, the larger the ratio
$\Gamma^0_{01}/\Gamma^0_{10}$ becomes. We remind that
$\Gamma^0_{01/10}$ is the transition rate from (to) the state with
$Q=k+1$ to (from) the state with $Q=k$. The increase of $\Delta$
makes the system less probable to be excited in the state with
$Q=k+1$ by the time-dependent gate voltage and, therefore, reduces
energy dissipation. Of course, this physical explanation is strongly
based on the model of only two charging states involved. It is valid
at $g\ll 1$. However, at $g^\prime(T)\gg 1$ the energy dissipation
rate has the maximum at half-integer values of the external charge
$q$ as well (see Eq.~\eqref{disone}). This result cannot be
explained by arguments based on the `orthodox' theory since there
are no well-defined charging states in the weak-coupling regime.

\begin{figure}%[h]
  \includegraphics[width=85mm]{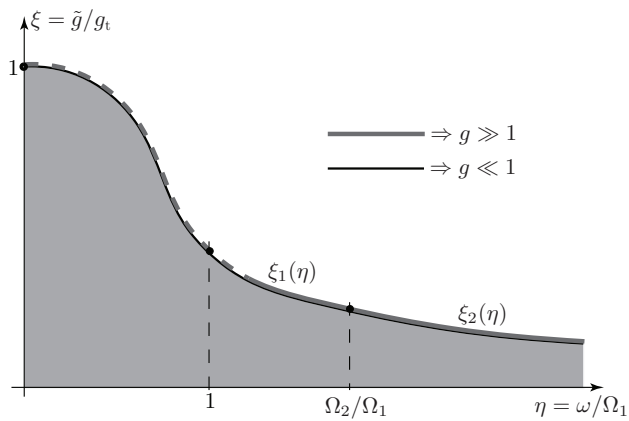}
  \caption{\label{figure9}  Schematic diagram for comparison of quantum and classical mechanisms of the energy dissipation. The quantum dissipation dominates in the filled region.
          }
\end{figure}

The dissipation caused by the electron tunneling is not the only one
that occurs in the set-up. Intrinsic electron transitions inside the
metallic island cause an additional internal energy loss. This
mechanism is, in fact, the origin of metallic conductivity.
%Since the conductivity is a classical quantity,
This sort of dissipation ought to be mainly a classical effect. It
corresponds to the radiation of energy by a metallic particle placed
in the quasi-stationary electric field. The classical dissipation
can be conveniently characterized by two limiting regimes: the
low-frequency ohmic loss and high frequency non-ohmic radiation
(skin-effect):
\begin{gather}
  \begin{split}
   W_\omega^c&\sim
   \frac{\hbar}{g_{\rm t}e^2}R^2\omega^2|U_\omega|^2,\quad\omega\ll\omega_0,\
   \ \omega_0=\frac{E_c}{g_{\rm t}\alpha^2\hbar}\\
   W_\omega^c&\sim\frac{\hbar}{g_{\rm t}e^2}R^2\omega^2\left(\frac{\omega}{\omega_0}\right)^{3/2}|U_\omega|^2,
   \quad\omega\gg\omega_0
  \end{split}
\end{gather}
Here,\ $\alpha=e^2/\hbar c$\ is a fine structure constant,\ $g_{\rm
t}e^2/\hbar$\ is an internal (Thouless) conductance of the island,\
$R$ - its characteristic size and\ $\omega_0$ is the separating
frequency. To elucidate the parametric conditions under which
quantum dissipation $\mathcal{W}_\omega$ due to presence of the
tunneling junction dominates over the classical one we make
necessary estimates. Quantum dissipation can also be split into
ohmic and non-ohmic limiting regimes. The corresponding separating
frequency is denoted as\ $\Omega$. We are concerned with simple
estimates only and drop weak log-corrections in all formulae for the
quantum case. The results are most transparently explained via phase
diagram which is presented in Fig.\ref{figure9} supplemented by
Tables~\ref{table1} and~\ref{table2}.
%%%%%%%%%%%%%%%%%%%%%%%%%%%%%%%%%%%%%%%%%%%%%%%%%%%%%%%%%%
\begin{table}
\caption{\label{table1}Description of functions for
Fig.~\ref{figure9}}
\begin{ruledtabular}
\begin{tabular}{c||c|c}
&\multicolumn{1}{c}{$\omega_0\leqslant \Omega$}&\multicolumn{1}{c}{$\omega_0> \Omega$}\\
  \hline
 $\Omega_1$ & $\omega_0$ & $\Omega$\\
 $\Omega_2$ & $\Omega$ & $\omega_0$\\
$\xi_1$ & $1/\eta^{3/2}$ & $1/\eta^2$\\
$\xi_2$ & $\frac{1}{\eta^{7/2}}\left(\frac{\Omega_2}{\Omega_1}\right)^2$ & $\frac{1}{\eta^{7/2}}\left(\frac{\Omega_2}{\Omega_1}\right)^{3/2}$\\
\end{tabular}
\end{ruledtabular}
\end{table}
%%%%%%%%%%%%%%%%%%%%%%%%%%%%%%%%%%%%%%%%%%%%%%%%
\begin{table}
\caption{\label{table2}Description of parameters for
Fig.~\ref{figure9}}
\begin{ruledtabular}
\begin{tabular}{c||cc}
 &\multicolumn{1}{c}{$\Omega$}&\multicolumn{1}{c}{$\tilde{g}$}\\
  \hline
  $g\gg1$                 & $gE_c/\hbar$ & $g$\\
  $g\ll 1$, $\Delta\ll T$ & $gT/\hbar$ & $g\left(\frac{T}{E_c}\right)^2$\\
  $g\ll 1$, $\Delta\gg T$ &$g\Delta/\hbar$ & $g\frac{\Delta T}{E_c^2}e^{\Delta/T}$\\
\end{tabular}
\end{ruledtabular}
\end{table}

In the fully coherent case, the admittance of SEB was studied in
Ref.~[\onlinecite{buttiker0}] by means of the $S$-matrix formalism.
It was shown that the SEB admittance can be presented in accordance
with its classical appearance \eqref{admittance_cl} but the
definition of physical quantities comprising it becomes different.
In Ref.~[\onlinecite{buttiker0}], it was derived that the gate
capacitance\ $C_g$\ and the tunneling resistance $R$ should be
substituted by the mesoscopic capacitance\ $C_\mu$ and the charge
relaxation resistance\ $R_q$, respectively. However, according to
our results, although being applicable in the fully incoherent case,
the SEB admittance in the quasi-stationary regime involves two
capacitances: the effective capacitance $\partial Q/\partial U_0$
which controls the imaginary part of $\mathcal{G}(\omega)$ and the
renormalized capacitance $C_g(T)$ which together with $R_q(T)$
determines the temperature behavior of $\Re\mathcal{G}(\omega)$. It
is the effective capacitance that corresponds to the mesoscopic
capacitance $C_\mu$. The appearance of the effective capacitance
$\partial Q/\partial U_0$ in the imaginary part of the admittance is
dictated by conservation of charge via the Ward
identity~\eqref{SWId}. We expect that the SEB admittance should
involve two physically different capacitances in general. Recently,
the SEB admittance was studied with the help of the $S$-matrix
formalism in the incoherent case also.~\cite{buttiker1} In
particular, it was predicted that in the fully incoherent regime and
at low temperatures the charge-relaxation resistance $R_q=h/(g
e^2)$. It is at odds with our result that $R_q=h/(e^2 g^\prime(T))$
since at low temperatures $g^\prime(T)$ can be very different from
$g$ (see Eqs.~\eqref{quantities2} and \eqref{gPpf}). The  reason
behind this discrepancy is as follows. Coulomb interaction in
Ref.~[\onlinecite{buttiker1}] was accounted for on the level of
classical equations of motion only, which was the conservation of
charge. In the mean time quantum fluctuations of charge are
significant throughout all our analysis and there is no obvious
justification to take them negligible.

To summarize, we have studied the energy dissipation in a single
electron box due to a slowly oscillating gate voltage.  We focused
on the regime of not very low temperatures when electron coherence
can be neglected but quantum fluctuations of charge are strong due
to Coulomb interaction. We considered cases of weak and strong
coupling. In both cases we found that the energy dissipation rate is
given by the same expression involving two physical observables
$g^\prime(T)$ and $C_g(T)$. Our result for the energy dissipation
rate can be obtained from the SEB equivalent circuit if one
substitutes $g^\prime(T)$ and $C_g(T)$ for $g$ and $C_g$,
respectively. We strongly believe that the universal expression we
found for the energy dissipation rate is valid for an arbitrary
value of the tunneling conductance.

\acknowledgements

The authors are indebted to L. Glazman, Yu. Makhlin, A.M.M.
Pruisken, A. Semenov and M. Skvortsov for helpful discussions. The
research was funded by RFBR (Nos. 09-02-92474-MHKC, 06-02-16533),
the Council for grants of the Russian President (No. 4445.2007.2),
the Dynasty Foundation, the Program of RAS ‘‘Quantum Macrophysics’’,
and CRDF.

\numberwithin{equation}{subsection}
\section{Appendix\label{Sec:App}}
\subsection{Energy derivative\label{Sec:App:First}}
Here we relate dissipation in the system to various field
correlators in weak and strong coupling regime.
\paragraph{Weak Coupling,\ $g\gg1$\label{Sec:App:First:a}}
We want to express correlator \eqref{polar-0} through AES effective
phase\ $\phi(\tau)$. We want to be rigorous and introduce Keldysh
contour.
\begin{figure}[ht]
  \includegraphics[width=65mm]{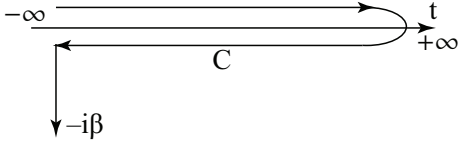}
   \caption{\label{figure10}
    Keldysh contour.
          }
\end{figure}
We split all the fields into upper and lower components\ ($\pm$)\ in
correspondence to the halves of the Keldysh contour. The action of
the system is split as well\ $S=S_{+}-S_{-}$\ and the partition
function of the system reads:\ ${\cal
Z}=\int\CD{\varphi_{\pm}}e^{iS[\varphi_{\pm}]}=1$. The average
electron density is found as:
\begin{gather}
   \label{density}
 \begin{split}
  \sum_\alpha\langle d_\alpha^\dagger d_\alpha\rangle&=
   \frac{1}{2}\sum_\alpha\Big\langle d^\dagger_{\alpha+}d_{\alpha+}+d^\dagger_{\alpha-}d_{\alpha-}\Big\rangle=\\
   &=\frac{C}{2C_g}\Big\langle\frac{\delta S}{\delta
   U_{g,q}}\Big\rangle+C_gU_c.
  \end{split}
\end{gather}
Here, we introduced classical and quantum components for bosonic
fields:
\begin{gather}
 U_{g,c}=\frac{1}{2}(U_{g+} +U_{g-}),\quad
 U_{g,q}=\frac{1}{2}(U_{g+}-U_{g-}),
\end{gather}
and\ $U_g(t)=U_0+U_\omega\cos\omega t$. To get rid of quartic
Coulomb terms we introduce Hubbard-Stratonovich boson fields:\
$V_+,\,V_-$\ on each part of the contour and make fermion gauge
transformation:
\begin{gather}
  \label{gauge_transform}
 d_{\alpha\sigma}\rightarrow
 d_{\alpha\sigma}e^{-i\int_0^tV_\sigma\,dt}.
\end{gather}
The transformed terms\ $S_0,\,S_c,\,S_t$\ take the form:
 \begin{gather}
 \begin{split}
   S_{0\sigma}&=\sum_k\int_{-\infty}^{\infty}
  a^\dagger_{k\sigma}\Big(i\partial_t-\varepsilon^{(a)}_k\Big)a_{k\sigma}\,dt\\
  &+\sum_\alpha\int_{-\infty}^{\infty}
   d^\dagger_{\alpha\sigma}\Big(i\partial_t-\varepsilon^{(d)}_\alpha\Big)d_{\alpha\sigma}\,dt,\\
  S_{c\sigma}&=\frac{C}{2}\int^\infty_{-\infty} V_\sigma^2\,dt+C_g\int^\infty_{-\infty} V_\sigma U_{g,\sigma}(t)\,dt,\\
  S_{t\sigma}&=-\sum_{\alpha,k}\int^\infty_{-\infty}\Big\{t_{k\alpha}e^{i\int V_\sigma dt}a_{k\sigma}^\dagger d_{\alpha\sigma}
  +\hbox{h.c.}\Big\}\,dt.
 \end{split}
\end{gather}
We see that the source term\ $U_{g,\sigma}$\ enter\ $S_{c\sigma}$\
only, hence we regroup it in a more suitable form:
\begin{gather}
 \label{density0}
\begin{split}
  S_c&=S_{c+}-S_{c-}=\\
  &=C\int_{-\infty}^\infty
  V_cV_q\,dt+\sqrt{2}C_g\int_{-\infty}^\infty\big(V_cU_{g,q}+V_qU_{g,c}\big)\,dt.
 \end{split}
\end{gather}
Here,\ $V_{c,q}=\frac{1}{\sqrt{2}}(V_+\pm V_-)$. Next we find the
physical electron density from (\ref{density}):
\begin{gather}
  \begin{split}
   \label{density1}
    \sum_\alpha\langle d_\alpha^\dagger d_\alpha\rangle&=\frac{C}{\sqrt{2}}\langle
    V_c\rangle+C_gU_{g,c}\ ,\\
    \bigg\langle\frac{\partial H}{\partial U_g}\bigg\rangle&=-\frac{C_g }{\sqrt{2}}\langle
    V_c\rangle.
   \end{split}
\end{gather}
We then expand\ $e^{iS}$\ to linear order in {\it physical} field\
$U_c$. The result reads
\begin{gather}
\label{ham-deriv}
  \begin{split}
  &\Big\langle\frac{\partial H}{\partial U_g}\Big\rangle=
  -\frac{C_g^2}{C^2}\int^\infty_{-\infty} \Pi_R(t-t^\prime)U_g(t^\prime)\,d t^\prime,
  \end{split}
\end{gather}
where\ $\Pi_R(t-t^\prime)=iC^2\langle V_c(t)V_q(t^\prime)\rangle$.
Coupled with (\ref{dissipation4}) it gives (\ref{dissipation2}).

Using \eqref{density1} we also write down the formula for the
effective capacitance\ $\partial Q/\partial U_0$\ of the SEB:
\begin{gather}
  \label{capacitance1}
   \frac{\partial Q}{\partial U_0}=C_g+\frac{\Pi_R(0)}{C}.
\end{gather}

\paragraph{Strong coupling,\ $g\ll1$\label{Sec:App:First:b}} We proceed in complete analogy with the
previous case. Using Hamiltonian (\ref{ham4}) we obtain
\begin{gather}
  \label{spin1}
   \bigg\langle\frac{\partial H}{\partial U_g}\bigg\rangle=-\frac{C_g}{C}
   \langle S^z(t)\rangle.
\end{gather}
Keldysh technique gives
\begin{gather}
   \langle S^z(t)\rangle=
   \frac{C_g}{C}\int_{-\infty}^{\infty}i\langle S_c^z(t)S_q^z(t^\prime)\rangle
   U_c(t^\prime)dt^\prime.
\end{gather}
Introducing spin-correlation function
\begin{gather}
   \label{spin2}
   \Pi^R_{s}(t)=i\langle S_c^z(t)S_q^z(0)\rangle
\end{gather}
we recover dissipation expression (\ref{spin-dissipation})\ with
spin correlator\ $\Pi^R_s(\omega)$\ playing the role of polarization
operator.
\subsection{Admittance}
The admittance is defined as:
\begin{gather}
   \label{admittance}
    \frac{\delta\langle I(t)\rangle}{\delta U_g(t^\prime)} =
    \int_{-\infty}^\infty {\cal
    G}(\omega)e^{-i\omega(t-t^\prime)}\frac{d\omega}{2\pi}.
\end{gather}
We introduce the tunneling current operator using hamiltonian\
(\ref{ham1}-\ref{ham3}):
\begin{gather}
    I=i[H,\sum_\alpha d_\alpha^\dagger d_\alpha]=i\sum_{k,\alpha}t_{k\alpha} a_k^\dagger
    d_\alpha+{\rm h.c.}
\end{gather}
To find the average current we insert the necessary source term into
the action:
\begin{gather}
   \label{source}
    S_s=\frac{1}{2}\int_{-\infty}^\infty I(t)\kappa(t)dt,\ \
    \langle I(t)\rangle=\frac{1}{i}\frac{\partial
    {\cal Z}[I]}{\partial\kappa(t)}\Big|_{\kappa=0}.
\end{gather}
While taking a functional integral along Keldysh contour we keep a
quantum component of\ $\kappa(t)$\ field only. We make the usual
rotation in the fermion basis:
\begin{gather}
  \begin{split}
     \psi_{\pm}&=\frac{1}{\sqrt{2}}(\psi_1\pm\psi_2),\\
     \bar{\psi}_{\pm}&=\frac{1}{\sqrt{2}}(\bar{\psi}_2\pm\bar{\psi}_1).
  \end{split}
\end{gather}
Here\ $\psi=(a_k,d_\alpha)^T$. After the rotation and gauge
transformation (\ref{gauge_transform}) the source and tunneling
terms take the form:
\begin{gather}
   \label{keldysh1}
   \begin{split}
    S_t+S_s&=\int dt\bar{\psi}_\gamma\Big( T_{\gamma\delta}(\varphi)+
    \frac{\kappa}{2}J_{\gamma\delta}(\varphi)\Big)\psi_\delta\\
    T_{\gamma\delta}&=\begin{pmatrix}
                          \Lambda_c\ &\ \Lambda_q\\
                          \Lambda_q\ &\ \Lambda_c
                       \end{pmatrix},\ \
   J_{\gamma\delta}=\begin{pmatrix}
                          J_q\ &\ J_c\\
                          J_c\ &\ J_q
                       \end{pmatrix}.\\
   \end{split}
\end{gather}
Here, indices\ $c$\ and\ $q$\ denote classical and quantum component
of a corresponding physical value, i.e.\
$J_{c,q}=\frac{1}{2}(J_{+}\pm J_{-})$\  and\ $\Lambda_\sigma,\
J_\sigma$\ are matrices in a island-lead space:
\begin{gather}
      \begin{split}
         \Lambda_\sigma=-
        \begin{pmatrix}
           0 &\ t_{k\alpha}e^{-i\varphi_\sigma}\\
           t^\dagger_{\alpha k}e^{i\varphi_\sigma} &\ 0
        \end{pmatrix},\\
         J_\sigma=
        \begin{pmatrix}
           0 &\ it_{k\alpha}e^{-i\varphi_\sigma}\\
           -it^\dagger_{\alpha k}e^{i\varphi_\sigma} &\ 0
        \end{pmatrix}.
      \end{split}
\end{gather}
It's possible to get rid of highly non-linear source term
(\ref{source}) by a suitable change of field variables. Indeed, one
can easily check that up to linear order in\ $\kappa$:
\begin{gather}
   T_{11}(\varphi_+,\varphi_-)+\frac{\kappa}{2}J_{11}(\varphi_+,\varphi_-)=
   T_{11}\Big(\varphi_++\frac{\kappa}{2},\varphi_--\frac{\kappa}{2}\Big).
\end{gather}
The same property holds for all the elements of matrices\
$T_{\gamma\delta},\ J_{\gamma\delta}$. By making a change:
\begin{gather}
   \varphi_++\frac{\kappa}{2}\rightarrow\varphi_+,\ \
   \varphi_--\frac{\kappa}{2}\rightarrow\varphi_-,
\end{gather}
we put the whole\ $\kappa$\ -dependence into gaussian part of the
action. Then
\begin{gather}
   S_s=-\int\kappa(t)(\frac{C}{\sqrt{2}}\ddot{\varphi}_c+C_g\dot{U}_c)dt.
\end{gather}
The average current (\ref{source}) reads
\begin{gather}
  \langle I\rangle=\frac{C}{\sqrt{2}}\langle\ddot{\varphi}_c\rangle+C_g\dot{U}_c\ .
\end{gather}
Using (\ref{density0})\ and to linear order in\ $U_c(t)$\ we find
the current to be
\begin{gather}
   \langle I\rangle=C_g\dot{U}_c+\frac{C}{\sqrt{2}}\langle\ddot{\varphi}_c\rangle+iCC_g
   \int U_c(t^\prime)dt^\prime\langle\ddot{\varphi}_c(t)\dot{\varphi}_q(t^\prime)\rangle.
\end{gather}
The admittance becomes
\begin{gather}
  \label{admittance_f}
  {\cal G}(\omega)=-i\omega C_g\Big(1+\frac{\Pi_R(\omega)}{C}\Big).
\end{gather}
Hence,
\begin{gather}
  \label{admittance1}
  \hbox{Im} \Pi_R(\omega)=\frac{C}{C_g}\frac{\hbox{Re}\, {\cal G}(\omega)}{\omega}.
\end{gather}
In the case of spin variables (strong coupling) we can easily get
the analogue of formula\ \eqref{admittance_f}\ for the admittance
using the same steps. This way we establish the relation between
admittance and spin polarization operator\ $\Pi_s$\ quoted in the
main body:
\begin{gather}
   \label{admittance_spin}
   {\cal G}(\omega)=-i\omega\frac{C_g}{C}\Pi^R_{s}(\omega)
\end{gather}
where\ $\Pi^R_{s}(\omega)$\ is given by\ \eqref{spin2}

\subsection{Instanton contributions\label{Sec:App:Second}}

\subsubsection{Massive fluctuations}
We expand the fluctuating field\ $\delta\varphi(\tau)$\ in the basis
of eigenfunctions:\ $\delta\varphi(\tau)=\sum_m C_m\varphi_m(\tau)$,
where the basis reads $(u=e^{2\pi iT\tau})$\,\cite{glazman}
\begin{gather}
  \label{basis}
  \begin{split}
  \varphi_m(\tau,z)&=u^{m-1}\frac{u-z}{1-u\bar{z}},\ \ m\ge2\ \
  ,\\
  \varphi_{-m}(\tau,z)&=\frac{1}{u^{m-1}}\frac{1-u\bar{z}}{u-z},\ \ m\ge2;\\
  \varphi_1(\tau,z)&=\sqrt{1-|z|^2}\frac{1}{u-z},\\
  \varphi_{-1}(\tau,z)&=\sqrt{1-|z|^2}\frac{u}{1-u{\bar z}}.
  \end{split}
\end{gather}
Here, $\varphi_{\pm1}(\tau,z)$ are field zero modes.  Then the
correlator reads
\begin{align*}
  &\langle{\cal T}_\tau\delta\dot{\varphi}(\tau)\delta\dot{\varphi}(\tau^\prime)\rangle
  =T\sum_{m}\int\hbox{\tencal D} z\dot{\varphi}_{-m}(\tau,z)\dot{\varphi}_{m}(\tau^\prime,z)\\
  &\times\langle C_{-m}C_m\rangle\frac{{\cal D}_1}{{\cal D}_0}e^{-g/2+2\pi iqW},\\
  &\langle C_{-m}C_m\rangle=\frac{2\pi}{g\omega_{m-1}}=\frac{1}{g(m-1)T},\ \ m>0\\
  &\hbox{\tencal D} z=\frac{d^2z}{1-|z|^2},\ \ \ |z|\leq1-\frac{T}{E_c}.
\end{align*}
Here,\ ${\cal D}_1/{\cal D}_0$\ is the ratio of fluctuation
determinants. Some care should be taken when regularizing them. We
used the scheme proposed in \cite{beloborodov1}
\begin{gather}
  \label{fluct-ratio}
    \frac{{\cal D}_1}{{\cal D}_0}=\frac{g^2E_c}{2\pi^3T}.
\end{gather}
After simple algebra we obtain
\begin{gather*}
 \begin{split}
  &\frac{1}{2gE_c T}e^{g/2-2\pi iqW}\langle{\cal T}_\tau\delta\dot{\varphi}(\tau)\delta\dot{\varphi}(\tau^\prime)\rangle_{W}=\\
  =&\underbrace{\frac{s}{(1-s)^2}\ln\frac{E_c}{T}}_{\hbox{\bf I}}-
  \underbrace{\frac{1}{s}\ln^2(1-s)}_{\hbox{\bf II}}-\underbrace{\frac{2s}{1-s}}_{\hbox{\bf III}}-\underbrace{\frac{2\ln(1-s)}{1-s}}_{\hbox{\bf IV}}+\\
  +&\Big(s\rightarrow\frac{1}{s}\Big),\\
  s&=\frac{u}{u^\prime}=e^{2\pi iT(\tau-\tau^\prime)}.
 \end{split}
\end{gather*}
 Expanding this expression into Teylor series
over\ $s$\ we get
\begin{align*}
  \hbox{\bf I}&=\sum_{n=1}^{\infty}ns^n\ln\frac{E_c}{T},\\
  \hbox{\bf II}&=2\sum_{n=1}^{\infty}\frac{s^n}{1+n}\sum_{k=1}^{n}\frac{1}{k}=2\sum_{n=1}^{\infty}\frac{H_n}{1+n}s^n,\\
  \hbox{\bf III}&=2\sum_{n=1}^{\infty}s^n,\\
  \hbox{\bf IV}&=-2\sum_{n=1}^{\infty}s^n\sum_{k=1}^{n}\frac{1}{k}=-2\sum_{n=1}^{\infty}H_n s^n.\\
\end{align*}
Here, $H_n$ is harmonic number. The contribution of gaussian
fluctuations into the correlator becomes
\begin{gather}
 \begin{split}
  &\frac{1}{2gE_cT}e^{g/2-2\pi iqW}\langle{\cal T}_\tau\delta\dot{\varphi}(\tau)\delta\dot{\varphi}(\tau^\prime)\rangle_W=\\
  =&\sum_{n=1}^{\infty}n\Big(\ln\frac{E_c}{T}-\frac{2H_n}{1+n}\Big)s^n-2\sum_{n=1}^{\infty}s^n
  +\Big(s\rightarrow\frac{1}{s}\Big).
 \end{split}
\end{gather}
Now we make analytical continuation of Fourier-components into the
region\ $n\ll1$. We are interested in linear in\ $n$\ term.
$$
  H_n=\frac{\pi^2n}{6}+{\cal O}(n^2).
$$
Extracting linear part and  summing instanton and anti-instanton
terms we obtain
\begin{gather}
 \begin{split}
  \langle{\cal T}_\tau\delta\dot{\varphi}(\tau)\delta\dot{\varphi}(\tau^\prime)\rangle_{n}&=-8gE_ce^{-g/2}
  \Big(1-\frac{|\omega_n|}{4\pi T}\ln\frac{E_c}{T}\Big)\cos2\pi q\\
  &+{\cal O}(\omega_n^2),
 \end{split}
\end{gather}
which does cancel partition function renormalization
\eqref{partition1}.
\subsubsection{Zero modes}
The corresponding single instanton configuration reads:
\begin{gather}
   \dot{\varphi}_W=2\pi TW\Big(\frac{u}{u-z}+\frac{{\bar z}u}{1-{\bar
   z}u}\Big),\ \ W=\pm1.
\end{gather}
The correlator is given by
\begin{gather}
  \begin{split}
  &\langle{\cal T}_\tau\dot{\varphi}(\tau)\dot{\varphi}(\tau^\prime)\rangle_{W}=e^{-g/2+2\pi
  iqW}(2\pi T)^2\frac{{\cal D}_1}{{\cal
  D}_0}\int\frac{d^2z}{1-|z|^2}\\
  &\times\sum_n\Big\{\big|z^2\big|^ns^n+\big|z^2\big|^ns^{-n}\Big\}.
  \end{split}
\end{gather}
The corresponding Fourier-component is as follows
\begin{gather}
  \langle{\cal T}_\tau\dot{\varphi}(\tau)\dot{\varphi}(\tau^\prime)\rangle_{n}=
  e^{-g/2}8\pi^2T\cos2\pi q\frac{{\cal D}_1}{{\cal D}_0}
  \int\frac{\big|z^2\big|^{|n|}}{1-|z|^2}\,d^2z.
\end{gather}
Expanding it in\ $n\ll1$\ to linear order we reproduce
\eqref{phi-phi-f}. %%
\subsection{Computation of polarization operator}
\subsubsection{The lowest order}
First we notice that\ $\hbox{I}(-\omega_n)=\hbox{II}(\omega_n)$.
Thus we will drop any odd function of\ $\omega_n$\ while calculating
I($\omega_n$). The analytical expression for diagram I (see
Fig.\ref{figure5}) reads
\begin{gather*}
 \begin{split}
  \hbox{I}(\omega_n)&=\frac{gT^2}{4\pi}\sum_{k,m}\frac{|\Omega_m|}{(i(\varepsilon_k+i\omega_n)-\xi_\sigma)^2}\\
  &\times\frac{1}{i(\varepsilon_k+\omega_n+\Omega_m)-\xi_{-\sigma}}\frac{1}{i\varepsilon_k-\xi_{\sigma}}.
 \end{split}
\end{gather*}
Performing the sum over fermion frequencies we get
\begin{gather*}
 \begin{split}
  &\hbox{I}(\omega_n)=\frac{gT}{4\pi}\\
  &\times\sum_{m}|\Omega_m|\bigg\{
    \frac{n_f(\xi_\sigma)}{(i\omega_n)^2}\bigg[
    \frac{1}{\Delta\sigma+ i(\omega_n+\Omega_m)}-\frac{1}{\Delta\sigma+\Omega_m}\bigg]\\
  &-\frac{n_f(\xi_{-\sigma})}{\Delta\sigma+i(\omega_n+\Omega_m)}\frac{1}{(\Delta\sigma+i\Omega_m)^2}
                  \bigg\}.
 \end{split}
\end{gather*}
where\ $n_f(x)=1/(e^{\beta x}+1)$\ is Fermi distribution function.
Simple algebra shows that\
$\hbox{I}(\omega_n)+\hbox{II}(\omega_n)=-\hbox{III}(\omega_n)$.
Thus, taking the limit\ $\eta\rightarrow-\infty$\ we obtain
\begin{gather*}
 \begin{split}
  &\hbox{I}(\omega_n)+\hbox{II}(\omega_n)+\hbox{III}(\omega_n)=-\frac{2gT}{\pi}e^{\beta\eta}\sinh\frac{\Delta}{2T}\sum_{m}|\Omega_m|\\
  &\times\bigg\{
    \frac{1}{(\Delta+i\Omega_m+i\omega_n)(\Delta+i\Omega_m)^2}+\omega_n\rightarrow-\omega_n
                  \bigg\}.
 \end{split}
\end{gather*}
Now it's clear that the sum over\ $\Omega_m$\ can be taken in terms
of digamma functions. The answer is given by Eq.(\ref{polar-1}).
%
%
%
%%%%%%%%%%%%%%%%%%%%%%%%%%%%%%%%%%%%%%%%%%%%%%%%%%%%%%%%%%%%%%%  Callan-Symanzik equation %%%%%%%%%%%%%%%%%%%%%%%%%%%%%%%%%%%%%%%%%%%%%%%%%%%%%%%%
%
%
%
\subsubsection{Callan-Symanzik equation for\ $\langle s_z\rangle$}
The anomalous dimension\ $\gamma$\ of operator\ $s^z_{pf}$\ is
introduced as
\begin{gather}
  \label{RG1}
  Z^\gamma s^{z,r}_{pf}(\bar{\Delta},\bar{g})=s^z_{pf}(\Delta,g,\Lambda).
\end{gather}
where\ $Z$\ is given by \eqref{green1_0} and\ $\Lambda$\ is a
cut-off:\ $\Lambda\sim E_c$. To extract $\gamma$ we write down the
corresponding CS-equation for the Green's function:\
$F_{pf}(\Delta,g,\Lambda)=\frac{1}{2}\sum_\sigma\sigma\langle\bar{\psi}_\sigma\psi_\sigma\rangle$.
The tree-level\ $F_{pf}$\ reads
\begin{gather}
  \label{RG2}
   F_{pf}(\Delta)=-e^{\beta\eta}\sinh\frac{\Delta}{2T}.
\end{gather}
Following general RG-philosophy and  with the help of (\ref{RG1}) we
write the corresponding CS-equation for function\
$F(\Delta,g,\Lambda)$\ in the form:
\begin{gather}
  \label{CS-equation}
  \Big(\frac{\partial}{\partial\ln \Lambda}+\beta_g\frac{\partial}{\partial g}+\beta_\Delta\frac{\partial}{\partial\Delta}-\gamma
  \frac{d\ln Z}{d\ln \Lambda}\Big)F_{pf}(g,\Delta,\Lambda)=0.
\end{gather}
where the corresponding\ $\beta$\ - functions are given by
\begin{gather}
  \label{RG}
  \begin{split}
   \beta_g=\frac{g^2}{2\pi^2},\quad
   \beta_\Delta=\frac{g\Delta}{2\pi^2}.
   \end{split}
\end{gather}
The term with\ $\beta_g$\ always contains extra\ $g$\ and can be
dropped in the leading order. Using action (\ref{pf-action}) we work
out the last term:
\begin{gather}
 \label{RG4}
  \frac{d\ln Z}{d\ln \Lambda}=-\frac{g}{4\pi^2}.
\end{gather}
To find\ $\gamma$\ we need to get\ $F$\ in the next to (\ref{RG2})
order:
\begin{gather}
  \label{RG3}
  \begin{split}
  F_{pf}(\Delta,g,\Lambda)&=-e^{\beta\eta}\sinh\frac{\Delta}{2T}\Big(1-\frac{g}{2\pi^2}\ln\frac{\Lambda}{\varepsilon}\Big)\\
  &+e^{\beta\eta}\frac{g\Delta}{4\pi^2T}\cosh\frac{\Delta}{2T}\ln\frac{\Lambda}{\varepsilon}.
  \end{split}
\end{gather}
Here,\ $\varepsilon$\ - is a characteristic scale of interaction.
Plugging (\ref{RG4})\ and\ (\ref{RG3}) into (\ref{CS-equation}) we
find:
\begin{gather}
  \gamma=2.
\end{gather}

\subsubsection{Exact expression for polarization operator}
In order to work out the polarization-operator diagram in
Fig.\ref{figure6} we follow the scheme proposed by Eliashberg
[\onlinecite{eliashberg}]. First we establish the analytical
properties of vertex function\
$\bm{\Gamma}(z,z+i\omega_n,i\omega_n)$\ as a function of complex
variable\ $z$. The operator expression for the vertex function reads
\begin{gather}
   \bm{\Gamma}_\sigma(\tau_1-\tau,\tau_2-\tau)=\langle {\cal T}_t\bar{\psi}_\sigma(t)\psi_\sigma(t)\bar{\psi}_{\sigma}(t_1)\psi_{\sigma}(t_2)\rangle.
\end{gather}
Its Lehman representation is as follows
\begin{gather}
  \label{lehman}
  \begin{split}
   &\bm{\Gamma}_\sigma(z,z+i\omega,i\omega)=T^4\sum_{nklm}W^\sigma_{lnmk}W^{*\sigma}_{lkmn}\\
   &\times\Bigg[
   \frac{e^{\beta\omega_{kn}}}{\omega_{kn}-i\omega}\Bigg\{\frac{e^{-\beta\omega_k}+e^{-\beta\omega_l}}{z+\omega_{kl}}-
   \frac{e^{-\omega_l}+e^{-\omega_n}}{z+i\omega-\omega_{ln}}\Bigg\}\\
   &+
   \frac{e^{\beta\omega_{lm}}}{\omega_{lm}-i\omega}\Bigg\{\frac{e^{-\beta\omega_n}+e^{-\beta\omega_m}}{z+\omega_{nm}}-
   \frac{e^{-\beta\omega_l}+e^{-\beta\omega_n}}{z+i\omega-\omega_{ln}}\Bigg\}\Bigg],\\
   &W^\sigma_{lnmk}=\langle l|\psi_{\sigma}|n\rangle\langle m|\psi_\sigma|k\rangle.
  \end{split}
\end{gather}
Complex calculus teaches us that the sum\ (\ref{lehman}) defines a
function with two horizontal cuts:\ $\hbox{Im}(z+i\omega)=0$\ and\
$\hbox{Im}(z)=0$. For simplicity let's restrict our attention to a
{\it retarded} vertex function\ $\omega_n>0$. Next we define three
vertex functions in accordance with the structure of cuts:
\begin{gather}
  \label{gamma-define}
   \begin{split}
       &\bm{\Gamma}^{RRR}(z,z+i\omega,i\omega)\quad{\rm if}\quad\hbox{Im} z>0,\\
       &\bm{\Gamma}^{ARR}(z,z+i\omega,i\omega)\quad{\rm if}\quad-i\omega_n<\hbox{Im} z<0,\\
       &\bm{\Gamma}^{AAR}(z,z+i\omega,i\omega)\quad{\rm if}\quad\hbox{Im}
       z<-i\omega_n.
   \end{split}
\end{gather}
The general expression for\ $\Pi_\sigma(i\omega_n)$\ then becomes:
\begin{gather}
 \begin{split}
   &\Pi_{\sigma}(i\omega_n)=\frac{T}{4}\sum_{\varepsilon_k}\\
   &\times\bm{\Gamma}_{\sigma}(i\varepsilon_k,i\varepsilon_k+i\omega_n,i\omega_n)G_{\sigma}(i\varepsilon_k+i\omega_n)
   G_{\sigma}(i\varepsilon_k)\\
   &=\oint_C
   \frac{d\varepsilon}{16\pi i}\tanh\frac{\varepsilon}{2T}
   \Gamma_{\sigma}(\varepsilon,\varepsilon+i\omega_n,i\omega_n)G_{\sigma}(\varepsilon+i\omega_n)G_{\sigma}(\varepsilon).
 \end{split}
\end{gather}
\begin{figure}[h]
  \includegraphics[width=60mm]{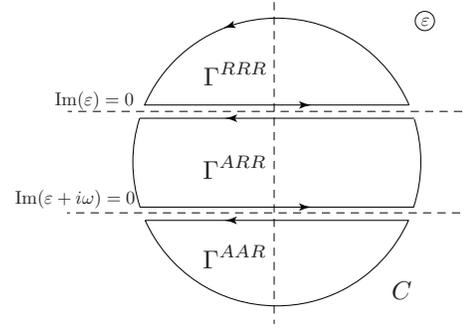}
   \caption{ Contour for polarization operator\ $\Pi(\omega)$.
   \label{figure11}
          }
\end{figure}
The contour\ $C$\ is shown in Fig.\ref{figure11}. As usual the
integral over large circle vanishes and we are left with integrals
over different branches:
%\begin{widetext}
\begin{gather}
 \begin{split}
   &\Pi_\sigma(i\omega_n)=\frac{1}{4}\int^{\infty}_{-\infty} \frac{d\varepsilon}{4\pi
   i}\tanh\frac{\varepsilon}{2T}\\
   \Big\{&\bm{\Gamma}^{RRR}_\sigma(\varepsilon,\varepsilon+i\omega_n,i\omega_n)G^R_\sigma(\varepsilon+i\omega_n)G^R_\sigma(\varepsilon)-\\
   -&\bm{\Gamma}^{ARR}_\sigma(\varepsilon,\varepsilon+i\omega_n,i\omega_n)G^R_\sigma(\varepsilon+i\omega_n)G^A_\sigma(\varepsilon)+\\
   +&\bm{\Gamma}^{ARR}_\sigma(\varepsilon-i\omega,\varepsilon,i\omega_n)G^R_\sigma(\varepsilon)G^A_\sigma(\varepsilon-i\omega_n)-\\
   -&\bm{\Gamma}^{AAR}_\sigma(\varepsilon-i\omega_n,\varepsilon,i\omega_n)G^A_\sigma(\varepsilon)G^A_\sigma(\varepsilon-i\omega_n)\Big\}.
 \end{split}
\end{gather}
%\end{widetext}
Making analytical continuation\ $i\omega_n\rightarrow\omega+i0$\ we
get result (\ref{polarization0}).
\subsubsection{Dyson equation for the vertex}
\begin{figure}
  \includegraphics[width=60mm]{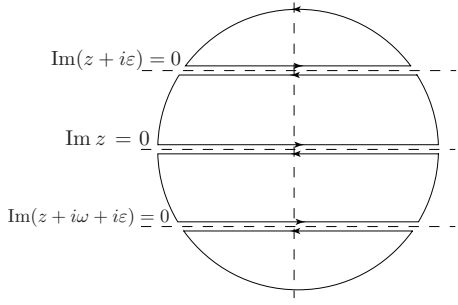}
   \caption{\label{figure12} Contour for the vertex function\ $\Gamma^{ARR}$.
          }
\end{figure}

Following the same scheme as in the previous section we derive the
expression for the vertex function. The contour\ $C$\ depends on the
type of the vertex we need to get from (\ref{vertex1}). The contour
for the vertex\ $\bm{\Gamma}^{ARR}_\sigma$\ is depicted in
Fig.\ref{figure12}.
The result reads
\begin{gather}
  \bm{\Gamma}^{ARR}_\sigma(\varepsilon,\varepsilon+\omega,\omega)=1-\hbox{I}+\hbox{II}-\hbox{III},
\end{gather}
where
%\begin{widetext}
\begin{gather}
  \begin{split}
    &\hbox{I}=\int_{-\infty}^{\infty}\frac{dx}{4\pi i}
         G^A_{-\sigma}(x)G^R_{-\sigma}(x+\omega)\bm{\Gamma}^{ARR}_{-\sigma}(x,x+\omega,\omega)\\
          &\times\Big[2i\hbox{Im}\,\alpha_R(x-\varepsilon)\coth\frac{x-\varepsilon}{2T}\\
         &-\tanh\frac{x}{2T}\alpha_R(x-\varepsilon)+\tanh\frac{x+\omega}{2T}\alpha_A(x-\varepsilon)\Big],\\
    &\hbox{II}=\int_{-\infty}^{\infty}\frac{dx}{4\pi i}\alpha_R(x-\varepsilon)
    G^R_{-\sigma}(x)G^R_{-\sigma}(x+\omega)\\
    &\times\bm{\Gamma}^{RRR}_{-\sigma}(x,x+\omega,\omega)
    \tanh\frac{x}{2T},\\
    &\hbox{III}=\int_{-\infty}^{\infty}\frac{dx}{4\pi i}\alpha_A(x-\varepsilon)
    G^A_{-\sigma}(x)G^A_{-\sigma}(x+\omega)\\
    &\times\bm{\Gamma}^{AAR}_{-\sigma}(x,x+\omega,\omega)\tanh\frac{x+\omega}{2T}.
  \end{split}
\end{gather}
%\end{widetext}
Here, function\ $\alpha(z)$\ is an interaction propagator whose
Matsubara counterpart is shown in Fig.\ref{figure4}. As usual it has
a cut\ $\hbox{Im}\, z=0$ which allows to define two functions:
\begin{gather}
  \alpha_R(\omega)=\bar{g}\frac{i\omega}{4\pi},\ \
  \alpha_A(\omega)=-\bar{g}\frac{i\omega}{4\pi}.
\end{gather}
The integrands entering terms II and III  are explicitly analytical
in the upper and lower halves of the complex plane respectively.
Consequently we may turn the corresponding integrals in to sums over
Matsubara frequencies\ $i\varepsilon_n$. Next one can easily prove
the following identities:
\begin{gather}
  \label{vertex3}
 \begin{split}
   \bm{\Gamma}_\sigma^{RRR}(i\varepsilon_n,i\varepsilon_n+\omega,\omega)&=
   \bm{\Gamma}_\sigma^{ARR}(i\varepsilon_n,i\varepsilon_n+\omega,\omega),\\
   \bm{\Gamma}_\sigma^{AAR}(i\varepsilon_n-\omega,i\varepsilon_n,\omega)&
   =\bm{\Gamma}_\sigma^{ARR}(i\varepsilon_n-\omega,i\varepsilon_n,\omega).
 \end{split}
\end{gather}
This way we drastically simplify our Dyson equation by rewriting it
entirely in terms of a single vertex\ $\bm{\Gamma}_{\sigma}^{ARR}$.
Next,
\begin{gather}
 \begin{split}
  \hbox{II}&=\frac{\bar{g}i}{4\pi}\sum_{\varepsilon_n}\bm{\Gamma}^{ARR}_{-\sigma}(i\varepsilon_n,i\varepsilon_n+\omega,\omega)\\
  &\times\frac{i\varepsilon_n-\varepsilon}{(i\varepsilon_n-\xi_{-\sigma})(i\varepsilon_n+\omega-\xi_{-\sigma})},\\
  \hbox{III}&=\frac{\bar{g}i}{4\pi}\sum_{\varepsilon_n}\bm{\Gamma}^{ARR}_{-\sigma}
  (i\varepsilon_n-\omega,i\varepsilon_n,\omega)\\
  &\times\frac{i\varepsilon_n-\omega-\varepsilon}{(i\varepsilon_n-\xi_{-\sigma})(i\varepsilon_n-\omega-\xi_{-\sigma})}.
 \end{split}
\end{gather}
As usual regularization scheme allows us to drop these sums. The
integrand of term I however contains\ $G^AG^R$. As a consequence it
is singular at\ $\omega,g\rightarrow0$\ as explained in the main
body. This way we recover Eq.\eqref{Dyson1_0}.
\subsection{Rate probabilities}
To work out rates\ $\Gamma^0$\ and\ $\gamma$\ we follow standard
scheme. We introduce Heisenberg $\psi$-operators according to
\begin{gather}
     \psi_d(t)=\sum_\alpha d_\alpha e^{-i\varepsilon_\alpha t},\quad
     \psi_a(t)=\sum_k a_k e^{-i\varepsilon_k t}.
\end{gather}
Then the matrix elements in the basis of filling numbers become
\begin{gather}
  \label{operator}
   \langle0|\psi_d|1\rangle=\sum_\alpha\langle0|d_\alpha|1\rangle e^{-i(\varepsilon_\alpha+\Delta)t}.
\end{gather}
In an ordinary fashion we change the Hamiltonian by gauge
transformation of fermion fields (path-integral approach is
implied):
\begin{gather}
   \psi_d(t)\rightarrow\psi_d(t) e^{i\frac{C_g}{C}\int U(t)dt},
\end{gather}
where\ $U(t)=U_\omega\cos\omega t$. Now the whole\ $U(t)$-dependance
is transferred into the tunneling part of the Hamiltonian:
\begin{gather}
  \label{tunnel-ham}
   H_t=\sum_{k,\alpha}t_{k\alpha}a_k^\dagger d_\alpha e^{i\frac{C_g}{C}\int
   U(t)dt}+\hbox{h.c.}
\end{gather}
Let us compute rate\ $\Gamma_{10}(t)$. The initial and final states
read
\begin{gather}
 \begin{split}
  |i\rangle&=|k,N\rangle,\\
  |f\rangle&=d^\dagger_{\alpha}a_k|k,N\rangle.
 \end{split}
\end{gather}
Here,\ $k\,(N)$\ is the number of electrons in the island (lead). As
usual,\ $S$-matrix formalism gives the necessary amplitude in the
form:
\begin{gather}
   \label{rate-amp1}
 \begin{split}
  A_{10}(t)&=-i\langle f|\int_{-\infty}^{t}H_t(t)dt|i\rangle\\
  &=-i\langle i|a^\dagger_k d_{\alpha}\int_{-\infty}^{t}H_t(t)dt|i\rangle.
 \end{split}
\end{gather}
Now we substitute $ \int U(t)dt=(U_\omega/\omega)\sin\omega t$ and
tunneling Hamiltonian assumes the form:
\begin{gather}
  \label{tunnel-ham1}
   H_t=\sum_{k,\alpha}t_{k\alpha}a_k^\dagger d_{\alpha}\Big(1+\frac{iC_gU_\omega}{C\omega}\sin\omega t\Big)+\hbox{h.c.}
\end{gather}
The detailed-balance relations for probability rates read
\begin{gather}
  \label{balance}
  \begin{split}
   \Gamma^0_{01}(\Delta)&=\Gamma^0_{10}(-\Delta),\\
   \gamma_{01}(t,\Delta)&=-\gamma_{10}(t,-\Delta).
  \end{split}
\end{gather}
Plugging (\ref{tunnel-ham1}) into (\ref{rate-amp1}) and integrating
one gets the amplitude of transition\ $0\rightarrow1$
\begin{gather}
  \label{amplitude1}
  \begin{split}
  &A_{10}(t)=t^\dagger_{\alpha k}(1-n_{\alpha})n_k e^{i(\varepsilon_{\alpha}-\varepsilon_k+\Delta)t}\\
  &\times\Bigg\{
       \frac{1}{\varepsilon_{k}-\varepsilon_{\alpha}-\Delta+i0}-\frac{C_gU_\omega}{2C\omega}\\
       &\times\Bigg[
       \frac{e^{i\omega t}}{\varepsilon_{k}-\varepsilon_{k^\prime}-\Delta-\omega+i0}-\frac{e^{-i\omega
       t}}{\varepsilon_{k}-\varepsilon_{k^\prime}-\Delta+\omega+i0}\Bigg]
     \Bigg\}.
  \end{split}
\end{gather}
Squaring it, taking thermal average and integrating we get the full
expression for the probability in the linear response regime:
\begin{gather}
  \begin{split}
  &W_{10}(t)=\frac{g}{8\pi^2}\int\frac{sds}{e^{\beta s}-1}
     \Bigg\{
       \frac{e^{2\lambda t}}{(s-\Delta)^2+\lambda^2}-\frac{C_gU_\omega}{C\omega}\times\\
       &\frac{1}{s-\Delta+i0}\bigg(
       \frac{e^{-i\omega t}}{s-\Delta-\omega-i0}-\frac{e^{i\omega
       t}}{s-\Delta+\omega-i0}\bigg)    \Bigg\}+\hbox{c.c.}
   \end{split}
\end{gather}
where\ $g$\ is defined in \eqref{conductance-def1}. Now we find the
transition rate as a derivative of a transition probability\
$\Gamma_{10}(t)=dW_{10}(t)/dt$\ and the following expression for\
$\gamma s$:
\begin{gather}
  \label{gamma-small10}
  \begin{split}
   &\gamma_{10}(\omega)=-\frac{g}{2\pi}\int_{-\infty}^\infty\frac{ds}{2\pi
   i}\frac{s}{e^{\beta s}-1}\times\\
    &\Bigg[\frac{1}{s-\Delta+i0}
         \frac{1}{s-\Delta-\omega-i0}\\
         &-\frac{1}{s-\Delta-i0}\frac{1}{s-\Delta+\omega+i0}\Bigg].
   \end{split}
\end{gather}
The integrand converges very well in the complex plane and the
integral can be easily taken:
\begin{gather}
  \label{gamma-small10-1}
  \begin{split}
   &\gamma_{10}(\omega)=-\frac{g}{2\pi}
    \Bigg[\frac{1}{\omega+i0}\Big(\frac{\Delta+\omega}{e^{\beta(\Delta+\omega)}-1}-\frac{\Delta}{e^{\beta\Delta}-1}\Big)+\\
    &+T\sum_{n=1}^{\infty}\frac{i\omega_n}{i\omega_n-\Delta}\Big(
         \frac{1}{i\omega_n-\Delta-\omega}-
         \frac{1}{i\omega_n-\Delta+\omega}\Big)\Bigg].
   \end{split}
\end{gather}
Expressing the sum in terms of digamma functions we get
\eqref{gamma-small-complete}. With the help of
\eqref{gamma-small10-1} and \eqref{balance} one can establish the
following useful identity:
\begin{gather}
    \label{gamma-identity}
     \gamma_{10}(\omega)-\gamma_{01}(\omega)=\frac{g}{2\pi}.
\end{gather}

%%%%%%%%%%%%%%%%%%%%%%%%%%%%%%%%%%%%%%%%
%%%%%%%%%%%%%%%%%%%%%%%%%%%%%%%%%%%%%%%%
%%%%%%%%%%%%%%%%%%%%%%%%%%%%%%%%%%%%%%%%


\begin{thebibliography}{100}
%% Intro

\bibitem{zaikin} G.\,Sch\"{o}n, A.\,Zaikin, Phys. Rep. \textbf{198}, 237 (1990).

\bibitem{ZPhys} The special issue on single charge tunneling, Z. Phys.
B \textbf{85}, 317 (1991).

\bibitem{grabert} For a review, see \textit{Single Charge Tunneling}, ed. by H.\,Grabert
and M.H.\,Devoret (Plenum, New York, 1992).

\bibitem{blanter}
Y.\,Blanter, M\,B\"{u}ttiker, Phys. Rep. {\bf 336}, 1 (2000).

\bibitem{aleiner}
I.\,Aleiner, P.\,Brouwer, L.\,Glazman, Phys. Rep. \textbf{358},  309
(2002).

\bibitem{Glazman} For a review, see L.I.\,Glazman and M.\,Pustilnik in \textit{New Directions in Mesoscopic Physics
(Towards to Nanoscience}, eds. R.\,Fazio, G.\,F.\,Gantmakher and
Y.\,Imry (Kluwer, Dordrecht, 2003).

\bibitem{buttiker0}
M.\,B\"{u}ttiker, H.\,Thomas, A.\,Pretre, Phys. Lett. A
\textbf{180}, 364 (1993).

\bibitem{buttiker3}
M.\,B\"{u}ttiker, A.M.\,Martin, Phys. Rev. B {\bf 61}, 2737 (2000).

\bibitem{buttiker2}
S.E.\,Nigg, R.\,L\'{o}pez, and M.\,B\"{u}ttiker, Phys. Rev. Lett.
{\bf 97}, 206804 (2006).

\bibitem{buttiker1}
M.\,B\"{u}ttiker, S.E.\,Nigg Phys. Rev. B {\bf 77}, 085312 (2008).

\bibitem{gabelli}
J.\,Gabelli, G.\,Feve, J.M.\,Berroir, B.\,Placais et al., Science
\textbf{313}, 499 (2006).

\bibitem{delsing}
F.\,Persson, C.M.\,Wilson, M.\,Sandberg, G.\,Johansson, P.\,Delsing,
arXiv:0902.4316.

\bibitem{matveev}
K.A.\,Matveev, Sov. Phys. JETP {\bf 72}, 892 (1991).

\bibitem{Grabert0} H.\, Grabert, Physica B {\bf 194-196}, 1011 (1994); Phys. Rev. B {\bf 50}, 17364 (1994).

\bibitem{matveev1}
K.A.\,Matveev, Phy.Rev. B {\bf 51}, 1743 (1995).

\bibitem{Grabert1} X.\,Wang and H.\,Grabert, Phys. Rev. B
\textbf{53}, 12621 (1996).

\bibitem{Grabert2} G.\,G\"{o}ppert,
H.\,Grabert, N.V.\,Prokof’ev, and B.V.\,Svistunov, Phys. Rev. Lett.
{\bf 81}, 2324 (1998).

\bibitem{Mahan} G.\,Mahan,
{\it Many particle physics}. (Plenum, New York, 2000), 3rd ed.


\bibitem{beloborodov1}
I.S.\,Beloborodov, A.V.\,Andreev, and A.I.\,Larkin, Phys. Rev. B
{\bf 68}, 024204 (2003).

\bibitem{imry} Z\,Ringel, Y.\,Imry, O.\,Entin-Wohlman, Phys. Rev. B {\bf 78} 165304 (2008).

\bibitem{Park} Hee Chul Park and Kang-Hun Ahn, Phys. Rev. Lett. {\bf 101}, 116804 (2008).

\bibitem{beloborodov}
I.\,Beloborodov, K.\,Efetov, A.\,Altland, and F.\,Hekking, Phys.
Rev. B {\bf 63}, 115109 (2001).

\bibitem{efetov}
K.B.\,Efetov and A.\,Tschersich, Phys. Rev. B {\bf 67}, 174205
(2003).

\bibitem{ambegaokar} V.\,Ambegaokar, U.\,Eckern and G.\,Sch\"{o}n, Phys. Rev. Lett. {\bf48},
1745 (1982).

\bibitem{burmistrov1}
I.S.\,Burmistrov, A.M.M.\,Pruisken Phys. Rev. Lett. {\bf 101},
056801 (2008)

\bibitem{glazman} A.\,Altland, L.\,Glazman, A.\,Kamenev, and J.\,Meyer, Ann. of Phys. (N.Y)
{\bf 321}, 2566 (2006).


%%%%%%%%%%%%%%%%%%%%%
%Sec2
%%%%%%%%%%%%%%%%%%%%%

\bibitem{LLIII} L.D.\, Landau and E.M.\, Lifshitz, {\it Course in Theoretical
Physics} (Pergamon, Oxford, 1981), Vol. 3.

\bibitem{LLV} L.D.\, Landau and E.M.\, Lifshitz, {\it Course in Theoretical
Physics} (Pergamon, Oxford, 1981), Vol. 5.

\bibitem{AGD} A.A.\, Abrikosov, L.P.\,Gorkov, and I.E.\,Dzyaloshinski,
{\it Methods of Quantum Field Theory in Statistical Physics} (Dover,
New York, 1963).

\bibitem{hofstetter} W.\,Hofstetter and W.\,Zwerger, Phys. Rev. Lett.
\textbf{78}, 3737 (1997); Eur. Phys. J. B {\bf 5}, 751 (1998).

\bibitem{perturb} F.\,Guinea and G.\,Sch\"{o}n, Europhys. Lett.
\textbf{1}, 585 (1986); S.A.\,Bulgadaev, JETP Lett. \textbf{45}, 622
(1987).

\bibitem{korshunov}
S.E.\,Korshunov, Pis’ma Zh. Eksp. Teor. Fiz. {\bf 45}, 342 (1987)
[JETP Lett. {\bf 45}, 434 (1987)].

\bibitem{Bulgadaev} S.A.\, Bulgadaev, Phys. Lett. A
\textbf{125}, 299 (1987).

\bibitem{panyukov} S.V.\, Panyukov and A.D.\, Zaikin, Phys. Rev. Lett. \textbf{67}, 3168
(1991).


\bibitem{polyakov} A.M.\,Polyakov,
{\it Gauge fields and strings}, (Harwood Academic Publishers, Shur,
1987).
\bibitem{schon1}
E.\,Ben-Jacob, E.\,Mottola and G.\,Sch\"{o}n,
 Phys. Rev. Lett. {\bf 51}, 2064 (1983); C. Wallisser et al., Phys. Rev. B {\bf 66},
125314 (2002).

\bibitem{burmistrov2}
I.S.\,Burmistrov, A.M.M.\,Pruisken, to be published.



%%%%%%%%%%%%%%%%%%%%%%%%%
%% Part IV

\bibitem{kulik} I.O.\,Kulik and R.I.\,Shekhter, Zh. Eksp. Teor. Fiz. {\bf 68}, 623
(1975) [Sov. Phys. JETP {\bf 41}, 308 (1975)];  E. Ben-Jacob and
Y.Gefen, Phys. Lett. A {\bf 108}, 289 (1985); K.K. Likharev and A.B.
Zorin, J. Low Temp. Phys. {\bf 59}, 347 (1985); D.V. Averin and K.K.
Likharev, J. Low Temp. Phys. {\bf 62}, 345 (1986).

\bibitem{abrikosov}
A.A.\,Abrikosov, Physics {\bf 2}, 21 (1965).

\bibitem{larkin}
A.I.\,Larkin and V.I.\,Melnikov, Zh. Eksp. Teor. Fiz. {\bf 61} 1231
(1971) [Sov. Phys. JETP {\bf 34}, 656 (1972)].

\bibitem{Sachdev} S. Sachdev and J. Ye, Phys. Rev. Lett. {\bf 70}, 3339 (1993).

\bibitem{Si} L.\,Zhu and Q.\,Si, Phys. Rev. B \textbf{66}, 024426 (2002).

\bibitem{Demler} G.\,Zar\'{a}nd and E.\,Demler, Phys. Rev. B \textbf{66}, 024427 (2002).

\bibitem{schon} G.\,Sch\"{o}n\ Phys. Rev. B {\bf 32}, 4469 (1985).

\bibitem{eliashberg} G.M.\,Eliashberg, Zh. Eksp. Teor. Fiz. {\bf 41}, 1241 (1961) [Sov. Phys. JETP {\bf 14},
886 (1962)].

\bibitem{schoeller} H.\,Schoeller and G.\,Sch\"{o}n, Phys. Rev. B
\textbf{50}, 18436 (1994).

%%%%%%%%%%%

\bibitem{ImryBook} Y. Imry, {\it Introduction to Mesoscopic Physics} (Oxford University, New York, 1997).

\bibitem{BlanterReview} Ya.M. Blanter, cond-mat/0511478 (unpublished).

\bibitem{Lesovik} G. B. Lesovik and R. Loosen, JETP Lett. 65, 295 (1997).

\bibitem{Deblock} R.\,Deblock, E\,Onac, L\,Gurevich, L.P.\,Kouwenhoven, Science {\bf 301}, 203 (2003); E.\,Onac F.\,Balestro, B.\,Trauzettel,
C. F.\,Lodewijk and L.P.\,Kouwenhoven , Phys. Rev. Lett. {\bf 96},
026803 (2006).


%\bibitem{Cohen}M.H.\,Cohen and A.M.M.\,Pruisken, Phys. Rev. B {\bf 49}, 4593 (1994).
%\bibitem{averin} D.V.\,Averin\,K.K.\,Likharev, in {\it Mesoscopic phenomena in Solids} (Elsevier, Amsterdam, 1991), p.173











\end{thebibliography}
\end{document}